\documentclass[applsci,review,accept,oneauthor,pdflatex,10pt,a4paper]{mdpi} 

\usepackage{verbatim}
\usepackage{amssymb} 
\usepackage{amsmath}
\usepackage{tabu}
\usepackage{longtable}

\renewcommand\Re{\operatorname{Re}}
\renewcommand\Im{\operatorname{Im}}
\renewcommand{\U}{\,\mathcal U}
\newcommand{\la}{\langle}
\newcommand{\ra}{\rangle}
\newcommand{\rt}{\right}
\newcommand{\lf}{\left}
\newcommand{\Tr}{\operatorname{Tr}}

\renewcommand{\k}{\mathbf k}
\newcommand{\s}{\text{s}}

\newcommand{\ks}{{\mathbf k_{\text{s}}}}
\renewcommand{\i}{\text{in}}

\newcommand{\br}{\mathbf r}
\newcommand{\el}{\text{el}}
\newcommand{\Q}{\mathbf Q}
\newcommand{\NRXS}{\text{NRXS}} 
\newcommand{\RXS}{\text{RXS}} 
\newcommand{\PE}{\text{PE}} 

\newcommand{\symbols}[1]{%
\vspace{12pt}\noindent{\selectfont\textbf{List of symbols}\par\vspace{6pt}\noindent {\fontsize{9}{9}\selectfont #1}\par}}

\newcommand{\commentout}[1]{\ignorespaces}
\firstpage{1} 
\articlenumber{318}
\doinum{10.3390/app8030318}
\pubvolume{8}
\pubyear{2018}
\copyrightyear{2018}
\history{Received: 12 January 2018; Accepted: 14 February 2018; Published: 25 February 2018}

\Title{Imaging Electron Dynamics with Ultrashort Light Pulses: A Theory Perspective}

\Author{Daria Popova-Gorelova $^{1,2}$\orcidA{}}

\AuthorNames{Daria Popova-Gorelova}

\address{%
$^{1}$ \quad Center for Free-Electron Laser Science, Deutsches Elektronen-Synchrotron DESY, Notkestrasse 85, D-22607~Hamburg, Germany; daria.gorelova@desy.de\\
$^{2}$ \quad The Hamburg Centre for Ultrafast Imaging, Luruper Chaussee 149, D-22761 Hamburg, Germany
}

\abstract{A wide range of ultrafast phenomena in various atomic, molecular and condense matter systems is governed by electron dynamics. Therefore, the ability to image electronic motion in real space and real time would provide a deeper understanding of such processes and guide developments of tools to control them. Ultrashort light pulses, which can provide unprecedented time resolution approaching subfemtosecond time scale, are perspective to achieve real-time imaging of electron dynamics. This task is challenging not only from an experimental view, but also from a theory perspective, since standard theories describing light-matter interaction in a stationary regime can provide erroneous results in an ultrafast case as demonstrated by several theoretical studies. We~review the theoretical framework based on quantum electrodynamics, which has been shown to be necessary for an accurate description of time-resolved imaging of electron dynamics with ultrashort light pulses. We compare the results of theoretical studies of time-resolved nonresonant and resonant X-ray scattering, and time- and angle-resolved photoelectron spectroscopy and show that the corresponding time-resolved signals encode analogous information about electron dynamics. Thereby, the information about an electronic system provided by these time-resolved techniques is different from the information provided by their time-independent analogues. 
}

\keyword{ultrafast light-matter interaction; electron dynamics; time-resolved imaging}

\begin{document}

\section{Introduction}

Real-time imaging of electron dynamics is one of the most important and challenging tasks for modern ultrafast science \cite{KrauszRMP09, SmirnovaNature09, GoulielmakisNature10, SansoneNature10, HaesslerNature10, TzallasNature11, HockettNature11, YoungJPhB18}. Valence electron dynamics occurs on a subfemtosecond to few-femtosecond timescale (1 fs = 10$^{-15}$ s) and its real-time measurement requires attosecond temporal resolution (1 as = 10$^{-18}$ s). At the same time, \r Angstrom spatial resolution is required to access inter-atomic distances in molecular structures and solids. 

Such ultrahigh spatial and temporal resolutions could be achieved using X-ray free-electron lasers, which can generate hard-X-ray pulses with \r Angstrom wavelengths \cite{CorkumNature07, GaffneyScience07, ChapmanNature06, VrakkingNature12, LeoneNature14}. Free-electron laser sources are capable of producing ultrashort X-ray pulses of femtosecond \cite{EmmaNature10, McNeilNature10} and even subfemtosecond duration \cite{ZholentsPRL04, EmmaPRL04, SaldinPhysRevSTAB06, KumarAppSciences13, TanakaPRL13, PratPRL15, CarbajoBanff15, HuangPRL17}. A remarkable achievement was demonstrated recently by \mbox{Huang  et al.,} who~succeeded to generate single spikes of hard X-rays that are only 200 attoseconds long at Linac Coherent Light Source \cite{HuangPRL17}. (Sub-)femtosecond hard X-ray pulses can be applied for time-resolved imaging of electron dynamics either by means of nonresonant X-ray scattering \mbox{(NRXS) \cite{DixitPNAS12, ZamponiPNAS12, DixitPRA14}} or by means of resonant X-ray scattering (RXS) \cite{PopovaGorelovaPRB15_1, PopovaGorelovaPRB15_2}, which are different processes determined by distinct terms of the light-matter interaction Hamiltonian. 

Alternatively to using photons for imaging electron dynamics with X-ray scattering, one~can consider employing photoelectrons dislodged by ultrashort light pulses for this goal. Time- and angle-resolved photoelectron spectroscopy (TRARPES),  i.e., time- and energy-resolved molecular-frame photoelectron angular distributions, using ultrashort extreme ultraviolet (XUV) probe pulses inducing single-photon ionization has been proposed for imaging coherent electron dynamics in molecules \cite{MignoletPRA12, KusJPCA13, PerveauxJPCA14, Popova-GorelovaPRA16}. High-energy photoelectrons generated by XUV pulses, on the one~hand, provide \r Angstrom spatial resolution and, on the other~hand, allow for a less sophisticated interpretation than that in the case of low-energy photoelectrons, where multiple scattering effects can be substatial \cite{SpringerBookAMO}.

The goal of this article is to show that imaging a nonstationary electronic system by ultrashort light pulses is nontrivial not only from an experimental but also from a theoretical perspective. We~aim to demonstrate that a simple extrapolation of an observable of a stationary measurement to a time-dependent quantity to obtain a time-resolved signal does not work and can lead to erroneous results. Therefore, a careful analysis of time-dependent processes governing a nonstationary measurement is necessary for a correct interpretation of its outcome.

For a measurement of a stationary electronic system by means of X-ray scattering, its electron density is the central quantity that determines a scattering pattern. Thus, one may naively expect that time-resolved scattering patterns from a nonstationary electronic system depend on its time-dependent electron density. However, it has been demonstrated that this assumption is not correct and time-resolved scattering patterns are not connected to a time-dependent electron density due to a considerable contribution of inelastic (Compton-type) processes that cannot be physically \mbox{avoided~\cite{DixitPNAS12, DixitPRA14, PopovaGorelovaPRB15_1, PopovaGorelovaPRB15_2, ShaoPRA13}}. Only taking into account all transitions that can be induced by a broadband, ultrashort probe pulse, one can obtain correct time-resolved scattering patterns. Such an analysis can be accurately performed within the quantum electrodynamics (QED) framework. 

Analogously, one may assume that electron dynamics can be measured by means of time-dependent chemical shifts using time-resolved photoelectron spectroscopy. Since electron binding energies, which determine photoelectron spectra, depend on the local chemical environment, one would think that photoelectron peaks would shift following a time-dependent electron density. However, an analysis of a time-resolved photoelectron probability within the QED framework has shown that this expectation is also incorrect \cite{Popova-GorelovaPRA16}. Thus, an intuitive approach to interpret an outcome of a time-resolved measurement can lead to erroneous results. 

In this article, we review the theoretical framework based on the QED, which has been developed to describe the interaction of an ultrashort light pulse with a coherently evolving electronic system. In particular, we will concentrate on three techniques, which have been suggested to image electron dynamics in real space and real time, and analyzed within the QED framework, namely, time-resolved~NRXS, time-resolved RXS and TRARPES. We will show that results of the QED analyses of the corresponding time-resolved signals can be represented in a common way although these techniques relay on different processes. Time-resolved signals obtained by means of these techniques have a similar temporal dependence on the evolution of the electronic state of a system being probed, which does not coincide with the temporal dependence of its electron density. We will discuss that accounting for certain features characteristic to the corresponding processes, analogous time- and space-dependent quantities of a nonstationary electronic system are encoded in corresponding time- and momentum-resolved signals. A common procedure based on a Fourier analysis of these signals can be applied to disentangle these quantities.

This review article is organized as follows. We describe a coherently evolving electronic system from a quantum-mechanical prospective in Section \ref{SectionElectronicSystem}. In Section \ref{SectionInteraction}, we introduce the basics of the QED treatment of the interaction of an ultrashort light pulse with such an electronic system. We review the QED analysis of time-resolved NRXS from a nonstationary electronic system and its results in Section~\ref{SectionNRXS}. The theory and suggested applications of time-resolved RXS employed for imaging of nonstationary electronic systems are reviewed in Section~\ref{SectionRXS}. The TRARPES technique is considered in Section \ref{SectionPE}. We discuss a limitation on a probe-pulse duration for an appropriate time-resolved measurement in Section \ref{SectionFrozenDensity}. We summarize the review in Section \ref{SectionDiscussion}.

\section{Coherently Evolving Electronic System}
\label{SectionElectronicSystem}

Nowadays, it is possible to trigger and observe coherent electron dynamics in atoms \cite{GoulielmakisNature10, TzallasNature11}, molecules \cite{SmirnovaNature09, SansoneNature10, HaesslerNature10, CalegariScience14} and crystals \cite{PolliNature07, KawakamiPRL10, KuehnPRL10, KuehnPRB10, SchubertNature14}, and even control the outcome of a simple chemical reaction \cite{RanitovicPNAS14}. Sub-femtosecond timing synchronization between pump and probe pulses required for observation of such dynamics has been achieved \cite{GoulielmakisNature10, BenedickNature12}. It has been demonstrated that valence-electron wave packets can evolve with a high degree of coherence for much longer than 10 fs \cite{GoulielmakisNature10}.

Let us consider such coherently evolving electronic system that had been excited by a pump pulse into a coherent superposition of the electronic eigenstates at time $t_0$. Its time evolution is given by
\begin{equation}
|\Psi(t) \ra = \sum_I C_Ie^{-iE_I(t-t_0)} |\Phi_I\ra.\label{EqWavePacket}
\end{equation}

 Here, $|\Phi_I\ra$ are the eigenstates and $E_{I}$ are the corresponding eigenenergies of the many-body Hamiltonian of the electronic system in the absence of an X-ray field, $\hat H_{\text{m}}$. We use atomic units for this and the following expressions. Since the purpose of this article is to discuss techniques that would allow probing such electron dynamics in real space and real time, we do not concentrate on a specific pump process giving rise to $|\Psi(t) \ra$. We assume that the pump and probe pulses do not overlap in time that makes it possible to describe the probe and the pump steps separately \cite{SantraPRA11}. 

The density matrix of this nonstationary electronic system given by 
\begin{align}
\hat\rho^m(t) =& |\Psi(t)\ra\la\Psi(t)|\label{EqDenMatrix}\\
= &\sum_{I,K}C_IC_K^*e^{-i(E_I-E_K)(t-t_0)}|\Phi_I\ra\la\Phi_K|\nonumber
\end{align}
 contains time-dependent complex off-diagonal elements. The time-dependent electron density of the electronic system is given by the relation $\rho(\mathbf r,t) = \Tr[\hat\psi^\dagger(\mathbf r)\hat\psi(\mathbf r) \hat\rho^m(t)]$ resulting in
\begin{align}
\rho(\mathbf r,t)&=\Re\la\Psi(t)|\hat\psi^\dagger(\mathbf r)\hat\psi(\mathbf r)|\Psi(t)\ra.\label{EqElDensity}
\end{align}

\section{Interaction with a Probe Pulse within the QED Framework}
\label{SectionInteraction}

When the electronic system is probed by an electromagnetic pulse, the total Hamiltonian of the whole system, matter and light, can be written as \cite{Loudon}
\begin{align}
\hat H =& \hat H_0+\hat H_{\text{int}}\nonumber\\
 =& \hat H_{\text{m}}+\sum_{\mathbf{k},p}\omega_{\k}\hat a_{\k,p}^\dagger\hat a_{\k,p}+\hat H_{\text{int}},
\end{align}
 where $\hat a_{\k,p}^\dagger$ and $\hat a_{\k,p}$ are creation and annihilation operators of a photon in the $\k$, $p$ mode of the radiation field with energy $\omega_{\k}=|\k|c$, where $c$ is the speed of light. 
$\hat H_{\text{int}}$ is the minimal coupling interaction Hamiltonian between the matter and the electromagnetic field in Coulomb gauge
\begin{align}
\hat H_{\text{int}}&=\hat H_{\text{int}}^{(1)}+\hat H_{\text{int}}^{(2)}\nonumber\\
&=\frac1c\int d^3r\hat \psi^\dagger(\mathbf r)\lf(\hat{\mathbf A}(\mathbf r)\cdot\mathbf p\rt)\hat \psi(\mathbf r)+\frac{1}{2c^2}\int d^3 r\hat \psi^\dagger(\mathbf r)\hat{\mathbf A}^2(\mathbf r)\hat \psi(\mathbf r),\label{H_int}
\end{align}
 where $\hat{\mathbf A}$ is the vector potential operator of the electromagnetic field, $\mathbf p$ is the canonical momentum of an electron, $\hat \psi^\dagger$ and $\hat \psi$ are electron creation and annihilation field operators. Depending on the character and conditions of the interaction of the electronic system with the electromagnetic pulse, one of two terms in Equation~(\ref{H_int}) becomes relevant for the interaction process and the other one turns to be insignificant. TRARPES, which is governed by absorption,  is determined by Hamiltonian $\hat H_{\text{int}}^{(1)}$. Scattering driven by the second term in Equation~(\ref{H_int}) is dominant in the case of high-energy X-rays and, thus, $\hat H_{\text{int}}^{(2)}$ determines NRXS. However, the scattering cross section given by $\hat H_{\text{int}}^{(1)}$ becomes much larger than that given by $\hat H_{\text{int}}^{(2)}$ in the case of a resonant X-ray pulse and, thus, $\hat H_{\text{int}}^{(1)}$ determines RXS.

In the case of X-ray scattering, the signal is determined by the probability $P^{\text{XS}}(\k_{\s})$ to observe a scattered photon with momentum $\k_{\s}$, which differs from the incoming photon momenta. Since we consider energy-unresolved scattering patterns, the signal is given by the energy average of $P^{\text{XS}}(\k_{\s})$,  i.e., by the differential scattering probability (DSP),
\begin{equation}
\frac{dP^{\text{XS}}}{d\Omega} = \frac{V}{(2\pi c)^3}\int_{0}^{\infty}d\omega_{\k_\s}\omega_{\k_\s}^2P^{\text{XS}}(\k_{\s})\label{DSP},
\end{equation}
 where $V$ is the quantization volume.

Within the density-matrix formalism, the expectation value of some observable $\hat O$ in a state, represented by a density matrix $\hat \rho$, is given by the relation $\la \hat O \ra = \Tr[\hat\rho\hat O]$ \cite{Mandel}. Consequently, the~probability of X-ray scattering is connected to the operator $\hat O_{\mathbf k_s}$, which describes the observation of a photon in the scattering mode $\k_\s$ independently on its polarization,
\begin{equation}
P^{\text{XS}}(\mathbf k_s)=\Tr\lf [\hat\rho_f(t_f)\hat O_{\mathbf k_s}\rt],\label{ProbabXS}
\end{equation}
 where $\hat\rho_f(t_f)$ is the total density matrix of the electronic system and the electromagnetic field at time $t_f$ after the action of the probe pulse \cite{Mandel, Loudon, DixitPNAS12}. The operator $\hat O_{\mathbf k_s}$ is given by
\begin{equation}
\hat O_{\ks} =\sum_{p_\s}W(\omega_{\k_\s})\hat a^\dagger_{\ks,p_\s}\hat a_{\ks,p_\s}\label{Oks}.
\end{equation}

Here, the sum is over two possible polarizations of a scattered photon. It is assumed that a photon detector has some acceptance range represented by the function $W(\omega_{\k_\s})$.

In the case of TRARPES, the signal is directly connected to the probability to observe an electron with momentum $\Q_\el$ given by
\begin{equation}
P^{\PE}(\Q_\el)=\Tr\lf [\hat\rho_f(t_f)\hat O_{\Q_\el}\rt],\label{ProbabPE}
\end{equation}
 where the operator 
\begin{equation}
\hat O_{\Q_\el} =\sum_{\sigma}\hat c^\dagger_{\Q_\el,\sigma}\hat c_{\Q_\el,\sigma}\label{Oks}
\end{equation}
 describes the observation of a photoelectron with momentum $\Q_\el$. $\hat c^\dagger_{\Q_\el,\sigma}$ and $\hat c_{\Q_\el,\sigma}$ are creation and annihilation operators of an electron with momentum $\Q_\el$ and spin $\sigma$ \cite{Mandel, Loudon, Popova-GorelovaPRA16}. Since we consider spin-unresolved photoelectron spectra, the sum is over spin $\sigma$.

Let us now consider the total density matrix $\hat \rho_f(t_f)$ of the matter and the electromagnetic field, which is obtained by the propagation of the initial total density matrix $\hat\rho_0$ with the time evolution operator $\hat \U(t_f,t_0)$, $\hat \rho_f(t_f) = \lim_{t_f\to+\infty}\hat \U(t_f,t_0)\hat\rho_0\hat\U^\dagger(t_f,t_0)$. The initial density matrix is given by $\hat\rho_0 = \hat \rho^m(t_0) \otimes \hat \rho^X_0$, where $\hat \rho^X_0 = \sum_{\{n\},\{\widetilde n\}}\rho ^X_{\{n\},\{\widetilde n\}}|\{n\}\ra\la\{\widetilde n\}|$ is the initial density operator of the electromagnetic field with $\{n\}$ and $\{\widetilde n\}$ being complete sets that specify the number of photons in all initially occupied field modes with a distribution $\rho ^X_{\{n\},\{\widetilde n\}}$ \cite{Loudon, Mandel}. Thus, the total density matrix can be represented as
\begin{align}
\hat \rho_f(t_f) =\lim_{t_f\to+\infty}\sum_{\{n\},\{\widetilde n\}}\rho ^X_{\{n\},\{\widetilde n\}}|\Psi_{\{n\}},t_f\ra\la\Psi_{\{\widetilde n\}},t_f|\label{rho_total},
\end{align}
 where an appropriate wave function $|\Psi_{\{n\}},t_f\ra$, which is an entangled state of the electronic and photonic states, is substituted dependent on the process considered as shown in the next Sections.

\section{QED Description of Time-Resolved Nonresonant X-ray Scattering}
\label{SectionNRXS}

Nonresonant X-ray scattering (usually, the specification 'nonresonant' is omitted) is an established technique that is employed to reveal structural information about a sample. In the scattering process, an electronic system in an initial state $I$ interacting with an X-ray pulse absorbs and emits a photon, which leaves a system either in the same state $I$ or brings it to a different final state $F$. The former case means that the photon has been scattered elastically and its energy is equal to the energy of the incoming X-ray beam $\omega_\i$. In the latter case, inelastic scattering has taken place and a photon with energy $\omega_\i-(E_F-E_I)$ has been scattered. Thus, elastic and inelastic scattering events from a stationary system can be distinguished by the spectroscopy of a scattered photon. Contributions due to elastic scattering to a scattering pattern dominate over that due to inelastic scattering, since transition amplitudes of elastic scattering events sum up coherently. Thus, scattering patterns from a stationary object are determined by elastic scattering and, as a result, encode the electron density $\rho(\br)$ of the object via the relation
\begin{align}
\frac{dP^{\text{st}\NRXS}(\Q)}{d\Omega} \propto\lf|\int d^3 r\rho(\br)e^{i\Q\cdot\br} \rt|^2\label{EqNRXSst},
\end{align}
 where $\Q$ is the scattering vector \cite{WarrenBook}. Solving the phase problem, one can reconstruct the electron density with the spatial resolution determined by the wave length of the X-ray pulse.

Let us now consider the interaction of an X-ray pulse with a nonstationary electronic system in the state $|\Psi(t) \ra = \sum_I C_Ie^{-iE_I(t-t_0)} |\Phi_I\ra$. In References~\cite{JuvePRL13, SuominenPRL14}, where this problem has been studied, it was assumed that the time-dependent electron density of a nonstationary electron system at the time of measurement $t_p$, $\rho(\br,t_p)$, should be substituted for the electron density $\rho(\br)$ in Equation~(\ref{EqNRXSst}) to obtain time-resolved scattering patterns
\begin{align}
\frac{dP^{\NRXS}(\Q)}{d\Omega} \propto\lf|\int d^3 r\rho(\br,t_p)e^{i\Q\cdot\br} \rt|^2.\label{EqNRXSwrong}
\end{align}

However, the connection of scattering patterns to the electron density stem from elastic scattering. If one would try to consider elastic scattering from an electronic system in the state $|\Psi(t) \ra = \sum_I C_Ie^{-iE_I(t-t_0)} |\Phi_I\ra$, one would have to assume that the final state of a system $F$ is exactly the same superposition $|\Psi(t) \ra$ as before the interaction with the X-ray pulse. In addition to the fact that the probability of such a situation is extremely weak, it is impossible to spectroscopically distinguish this scattering event from other transitions with final states within the eigenstates of the wave packet $|\Psi(t) \ra$. In view of this, the concept of `elastic scattering' in the case of the interaction of an X-ray pulse with a nonstationary electronic system is ambiguous, and the connection of scattering patterns to the electron density must be carefully reconsidered for the time-dependent case \cite{DixitPNAS12}.

In order to accurately describe the interaction of an X-ray pulse with an electronic system evolving coherently, a theoretical analysis employing the QED framework has been performed by \mbox{Dixit  et al.} in Reference~\cite{DixitPNAS12}. Let us briefly review the result of this work. As discussed in the previous Section, nonresonant X-ray scattering is governed by the Hamiltonian $\hat H_{\text{int}}^{(2)}$ in Equation~(\ref{H_int}). In order to obtain the probability of nonresonant X-ray scattering with Equation~(\ref{ProbabXS}), the total density matrix $\hat \rho_f(t_f)$ must be evaluated within the first-order time-dependent perturbation theory using $\hat H_{\text{int}}^{(2)}$ as the perturbation. The resulting density matrix is connected to the first-order wave function
\begin{align}
\lf|\Psi_{\{n\}}^{\NRXS},t_f\rt\ra=&-i\int_{t_0}^{t_f}dt\,e^{i\hat H_0 t}\hat H_{\text{int}}^{(2)}\,e^{-i\hat H_0 t}|\{n\}\ra\lf|\sum_I C_I\Phi_I\rt\ra,\label{Psi1}
\end{align}
 which must be substituted for $|\Psi_{\{n\}},t_f\ra$ in Equation~(\ref{rho_total}). Thus, according to Equations~(\ref{DSP}) and (\ref{ProbabXS}), the~DSP of NRXS from an electronic system evolving coherently is

\begin{align}
\frac{dP^{\NRXS}(\ks)}{d\Omega} 
=&\frac{\sum_{p_\s} \bigl|(\boldsymbol\epsilon_{\i}\cdot \boldsymbol\epsilon^*_{s})\bigr|^2}{4\pi^2c^3\omega_{\i}^2}
\int_{0}^{\infty}d\omega_{\k_\s}\omega_{\k_\s} W(\omega_{\k_\s})\sum_{F} \int_{t_0}^{+\infty} dt_1\int_{t_0}^{+\infty} dt_2\label{EqDSP_NRXS_Gen}\\
&\qquad\times\int d^3 r_1\int d^3 r_2 G^{(1)}(\br_1,t_1,\br_2,t_2)e^{-i\ks\cdot(\br_1-\br_2)}M_{F}^{\NRXS}(\br_1,t_1)\bigl[M_{F}^{\NRXS}(\br_2,t_2)\bigr]^\dagger.\nonumber
\end{align}
 with the function
\begin{align}
M_{F}^{\NRXS}(\br,t) = &e^{i (E_F+\omega_{\ks}) t}\la \Phi_F |\hat\psi^\dagger(\br) \hat\psi(\br)|\Psi(t)\ra.
\end{align}
%

 Here, $G^{(1)}(\br_1,t_1,\br_2,t_2)=\Tr[\hat \rho^X_0\hat {\mathbf E}^{-}\hat {\mathbf E}^{+}]$ is the first-order radiation field correlation function, where $\hat {\mathbf E}^{+}=(\hat {\mathbf E}^{-})^\dagger=\sum_\k\sqrt{2\pi\omega_\k/V}\hat a_{\k,p}\boldsymbol\epsilon_p e^{-i\omega_\k t+i\k\cdot\br}$ with $\boldsymbol\epsilon_p$ being the unit vector corresponding to the polarization $p$ \cite{GlauberPhRev63, Loudon}. $\omega_\i$ is the photon energy and $\boldsymbol\epsilon_\i$ is the mean polarization vector of the incoming X-ray beam. The sum is over all possible final states $F$ with the corresponding energies $E_F$. The~role of the function $W(\omega_{\k_\s})$ is to represent the acceptance range of the photon detector, which~can, for~example, resolve only certain scattered energies $\omega_{\k_\s}$. This expression considers a general case, which is independent on a probe-pulse duration and coherence properties.

Equation~(\ref{EqDSP_NRXS_Gen}) can be simplified under certain conditions. Dixit  et al.~considered a perfectly coherent Gaussian-shaped probe pulse with the amplitude of the electric field
\begin{align}
\mathcal E(\mathbf r_0,t)=\sqrt{(8\pi/c) I_0(\mathbf r_0)}e^{-2\ln2\lf(\frac{t-t_p}{\tau_p}\rt)^2},\label{EqGaussianPulse}
\end{align}
 where $\mathbf r_0$ is the position of the object, $t_p$ is the time of the measurement, $\tau_p$ is the pulse duration (FWHM of the pulse intensity) and $I_0(\mathbf r_0)=cE^2(\mathbf r_0,t=0)/(8\pi)$ is the peak intensity. It was further assumed that the probe-pulse duration is much shorter than the characteristic time scale of changes in the electron density. Hereafter, we will refer to this assumption as the frozen-density approximation. As we discuss in Section \ref{SectionFrozenDensity}, this is a necessary condition for an appropriate time-resolved measurement. The resulting time-dependent DSP can be represented as
\begin{align}
\frac{dP^{\NRXS}(\Q)}{d\Omega} =\mathrm P_0^{\NRXS} \sum_F \int d^3 r_1\int d^3 r_2 e^{i\Q\cdot(\br_1-\br_2)}\la\Psi(t_p)|\hat G_F^\NRXS(\br_1,\br_2)|\Psi(t_p)\ra e^{-\Omega_F^2\tau_p^2/(4\ln2)}\label{EqDSP_NRXS}
\end{align}
 with the operator
\begin{align}
\hat G_F^\NRXS =
\hat \psi(\br_2) \hat \psi^\dagger(\br_2)|\Phi_F\ra\la\Phi_F|\hat \psi(\br_1) \hat \psi^\dagger(\br_1)\int_{0}^{\infty}d\omega_{\k_\s}\omega_{\k_\s} W(\omega_{\k_\s}).\label{EqGF_NRXS}
\end{align}

Here, $\mathbf Q = \k_\i-\ks$ is the scattering vector with $\k_\i$ being the mean wave vector of the incoming X-ray beam, $\mathrm P_0^{\NRXS} = I_0\tau_p^2\sum_{p_\s}\bigl|(\boldsymbol\epsilon_{\i}\cdot \boldsymbol\epsilon^*_{\s})\bigr|^2/(4\ln2\,\omega_{\i}^2c^3)$ and $\Omega_F = \omega_\i-\omega_{\k_\s}-E_F+\la E\ra$ with $\la E\ra$ being the mean energy of the coherent superposition in Equation~(\ref{EqWavePacket}).

Let us now compare the resulting relation for the time-resolved DSP in Equation~(\ref{EqDSP_NRXS}) with the relation in Equation~(\ref{EqNRXSwrong}). Equation~(\ref{EqNRXSwrong}) that can be represented as $dP^{\NRXS}(\Q)/d\Omega \propto \int d^3 r_1\int d^3 r_2 e^{i\Q\cdot(\br_1-\br_2)}\la\Psi(t_p)|\hat\psi^\dagger(\mathbf r_1)\hat\psi(\mathbf r_1)|\Psi(t_p)\ra \la\Psi(t_p)|\hat\psi^\dagger(\mathbf r_2)\hat\psi(\mathbf r_2)|\Psi(t_p)\ra$ assumes that elastic scattering has taken place and the final state coincides with the initial electronic state $|\Psi(t_p)\ra$. Equation~(\ref{EqDSP_NRXS}) contains a sum over final states $F$, which are eigenstates of the Hamiltonian of the electronic system $\hat H_{\text{m}}$, but not a superposition of electronic states. Contributions due to scattering to final states $F$ are summed incoherently and are weighted by the function $e^{-\Omega_F^2\tau_p^2/(4\ln2)}$ originating from the probe-pulse spectral density. Even if one could spectroscopically resolve scattering exclusively to final states, which are involved in the wave packet $\Psi(t_p)$, a contribution due to scattering to each eigenstate would have to be added incoherently, and it would not be possible to substitute $\Psi(t_p)$ for $\Phi_F$ in Equation~(\ref{EqGF_NRXS}). As a result, the state $\Psi(t_p)$ enters Equation~(\ref{EqDSP_NRXS}) twice instead of four times as in Equation~(\ref{EqNRXSwrong}). Thus, the assumption that time-resolved X-ray scattering is given the Fourier transform of the time-dependent electron density is incorrect and even fails to reflect the correct time-dependence of scattering patterns on the evolution of the electronic system.  

\begin{figure}[H]
\centering
\includegraphics[width=12 cm]{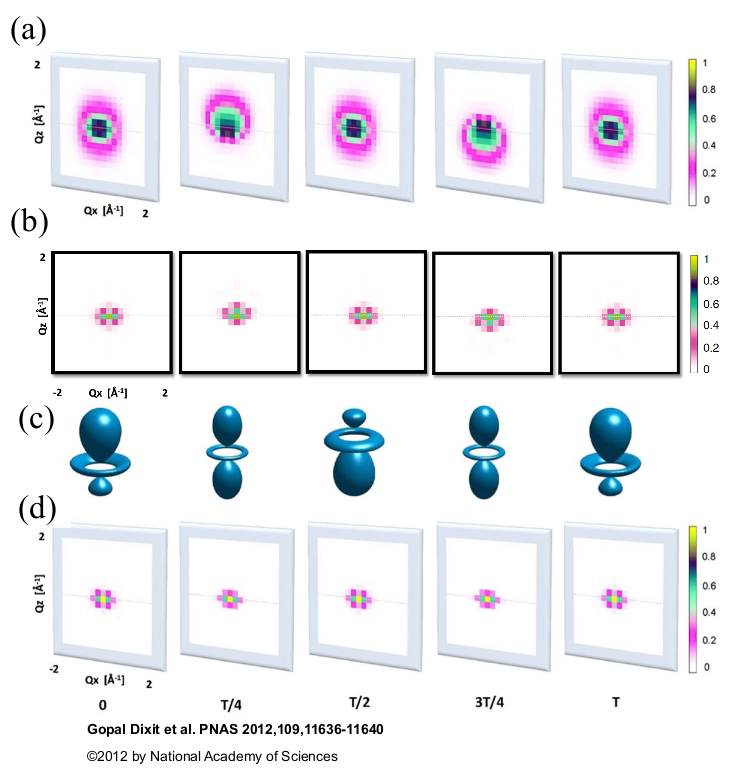}
\caption{Scattering patterns in the $Q_x-Q_z$ plane ($Q_y=0$) and electronic charge distributions of the wavepacket $\Psi(t)$. (\textbf{a}) Scattering patterns obtained with Equation~(\ref{EqDSP_NRXS}); {(\textbf{b}) scattering patterns obtained with Equation~(\ref{EqDSP_NRXS}), but the calculation is restricted to the two electronic states involved in the dynamics}; (\textbf{c}) electronic charge distributions and (\textbf{d}) scattering patterns obtained with Equation~(\ref{EqNRXSwrong}), at pump-probe delay times 0, $T/4$, $T/2$, $3T/4$, and $T$, where the oscillation period of the electronic wavepacket is $T = 6.25$ fs. The intensities of the patterns are shown in units of the DSP from a free electron in all cases. Figure is adopted with permission from Reference~\cite{DixitPNAS12}. Copyright National Academy of Sciences, 2012.
\label{PNASFig}
}
\end{figure}

In order to illustrate the difference between Equations~(\ref{EqNRXSwrong}) and (\ref{EqDSP_NRXS}), Dixit  et al.~considered an electronic wave packet prepared by a pump pulse as a coherent superposition $\Psi(t) = 1/\sqrt2 e^{-iE_{3d}(t-t_0)}\,|3d\ra+1/\sqrt2 e^{-iE_{4f}(t-t_0)}\,|4f\ra$ of the $3d$ and $4f$ eigenstates of atomic hydrogen with the projection of orbital angular momentum equal to zero, the polarization direction along the $z$-axis and the corresponding energies $E_{3d}$ and $E_{4f}$. The probe X-ray pulse was assumed to have duration of 1~fs, 4~keV photon energy and propagate along the $y$ direction. They calculated scattering patterns from the electronic wave packet with Equation~(\ref{EqNRXSwrong}) and with Equation~(\ref{EqDSP_NRXS}) and compared them in~Figure~\ref{PNASFig}.

Figure \ref{PNASFig}a shows scattering patterns calculated with Equation~(\ref{EqDSP_NRXS}) within the QED framework. Their structure is much more diverse than the structure of the semiclassical scattering patterns calculated with Equation~(\ref{EqNRXSwrong}) shown in Figure~\ref{PNASFig}d due to inelastic processes that have to be taken into account in Equation~(\ref{EqDSP_NRXS}). Namely, the calculation of the scattering patterns in Figure~\ref{PNASFig}a considered all possible final states matching the assumed photon detection width of 0.5 eV. However,~the most important difference appears at delay times $T/4$ and $3T/4$, when the electronic charge distributions are identical, but the electronic wave packet carry a different phase. The correct scattering patterns in Figure~\ref{PNASFig}a are different from each other at delay times $T/4$ and $3T/4$ unlike the scattering patterns at delay times $T/4$ and $3T/4$ in Figure~\ref{PNASFig}d, which are equal to each other. Thus, the correct scattering patterns depend on the phase of the wave packet, but not on the electron density at the time of measurement as the scattering patterns in Figure~\ref{PNASFig}d do. The further principal difference is that the scattering patterns in Figure~\ref{PNASFig}a are not centrosymmetric (i.e., they are not equal for $\Q$ and $-\Q$) at delay times $T/4$ and $3T/4$ breaking the Friedel's law \cite{Als-NielsenBook}.  For the sake of comparison, \mbox{Dixit {et al.}} have shown scattering patterns calculated with Equation~(\ref{EqDSP_NRXS}) within the QED framework, but~restricting the calculation to the two electronic states involved in the dynamics [see Figure~\ref{PNASFig}b]. The structure of these patterns is similar to the structure of the semiclassical patterns in Figure~\ref{PNASFig}d. However, analogously to the patterns in Figure~\ref{PNASFig}a, they~depend on the phase of the electronic wave packet and are not centrosymmetric at delay times $T/4$ and $3T/4$. Thus, not only inelastic transitions to a manifold of possible final states do lead to a disagreement between the QED and semiclassical scattering patterns, but also an incorrect description of the time dependence of the DSP in Equation~(\ref{EqNRXSwrong})~does. 

Obviously, the connection to the electronic state encoded in time-resolved scattering pattern is much less straightforward as suggested by Equation~(\ref{EqNRXSwrong}). In order to interpret information encoded in time-resolved scattering patterns, a Fourier analysis has been proposed for time-resolved resonant scattering patterns \cite{PopovaGorelovaPRB15_2}. Let us apply such an analysis to the nonresonant case. Equation~(\ref{EqDSP_NRXS}) can be represented as
\begin{align}
\frac{dP^{\NRXS}(\Q)}{d\Omega} =&\int d^3 r_1\int d^3 r_2 \cos[\Q\cdot(\br_1-\br_2)] \mathcal R^{\NRXS}(t_p,\br_1,\br_2)\label{EqDSP_NRXS_JR}\\
&-\int d^3 r_1\int d^3 r_2 \sin[\Q\cdot(\br_1-\br_2)] \mathcal J^{\NRXS}(t_p,\br_1,\br_2).\nonumber
\end{align}

 The function entering the expression above
\begin{align}
\mathcal R^{\NRXS}(t_p,\br_1,\br_2) =& \mathrm P_0^{\NRXS}  \Re\lf[\la\Psi(t_p)|\sum_F\hat G_F^{\NRXS}(\br_1,\br_2)|\Psi(t_p)\ra \rt]e^{-\Omega_F^2\tau_p^2/(4\ln2)}
\end{align}
 is connected to the real part of the elements of the electron density matrix $\hat\rho^m(t_p)$ [{cf.}~Equation~(\ref{EqDenMatrix})]. Therefore, its temporal dependence correlates with the electron density [{cf.}~Equation~(\ref{EqElDensity})] and related quantities such as charge distributions at the time $t_p$. Please notice that this temporal dependence is different to that given by the semiclassical Equation~(\ref{EqNRXSwrong}), which correlates with the time-dependent electron density squared. The corresponding term in Equation~(\ref{EqDSP_NRXS_JR}) is centrosymmetric with respect to~$\Q$. The function 
\begin{align}
\mathcal J^{\NRXS}(t_p,\br_1,\br_2) =& \mathrm P_0^{\NRXS}  \Im\lf[\la\Psi(t_p)|\sum_F\hat G_F^{\NRXS}(\br_1,\br_2)|\Psi(t_p)\ra \rt]e^{-\Omega_F^2\tau_p^2/(4\ln2)}
\end{align}
 is connected to the imaginary parts of the electron density matrix elements and its temporal dependence correlates with the electron current density, which is given by
\begin{align}
\mathbf j(\mathbf r, t_p) = \Im\lf[\la\Psi(t_p)| \hat\psi^\dagger(\br)\boldsymbol\nabla\hat\psi(\br)|\Psi(t_p)\ra\rt]\label{ProbCurrDen}.
\end{align}

The corresponding term in Equation~(\ref{EqDSP_NRXS_JR}) is an odd function with respect to $\Q$ and is responsible for the time-resolved scattering patterns being noncentrosymmetric. Thus, the fact that the scattering patterns in Figure~\ref{PNASFig} are not centrosymmetric reflects that the electron current is nonzero,  i.e., that~electrons are moving.

The two contributions to the scattering patterns can be decomposed performing the Fourier transform of the DSP $\mathcal F^{\NRXS}(\br,t_p ) = 1/(2\pi)^3\int d^3Q e^{-i\Q\cdot\br}dP^{\NRXS}(\Q)/d\Omega$. The real part of the Fourier transform is connected to the function $\mathcal R^{\NRXS}$ via the relation
\begin{align}
\Re\lf[ \mathcal F^{\NRXS}(\br,t_p)\rt] = \int d^3 r' \mathcal R^{\NRXS}(t_p,\br',\br'-\br),
\end{align}
and the imaginary part is connected to the function $\mathcal J^{\NRXS}$ according to
\begin{align}
\Im\lf[ \mathcal F^{\NRXS}(\br,t_p)\rt] = \int d^3 r' \mathcal J^{\NRXS}(t_p,\br',\br'-\br).
\end{align}

Thus, performing the Fourier transform of the scattering patterns from $\Q$ space to real space obtained at different time delays $t_p-t_0$, one obtains a complex function, which depends on the time of measurement $t_p$ and the space coordinates $\br$. The real part of this function correlates with charge distributions at the time of measurement $t_p$, and its imaginary part correlates with electron currents at~$t_p$. A~more specific connection of $\Re\lf[ \mathcal F^{\NRXS}(\br,t_p)\rt]$ and $\Im\lf[ \mathcal F^{\NRXS}(\br,t_p)\rt]$ to certain time-dependent quantities has to be identified for a given electronic system. Such an analysis has been performed for the case of RXS \cite{PopovaGorelovaPRB15_1, PopovaGorelovaPRB15_2} as described in the next Section.

Several alternative approaches to image electron dynamics by means of time-resolved NRXS have been developed through the application of the QED framework. Dixit  et al. has proposed to image the instantaneous electron density of an electronic wave packet via ultrafast X-ray phase contrast imaging \cite{DixitPRL13}. They have shown that ultrafast phase contrast imaging provides the Laplacian of the electron density, which reveals complex bonding and topology of the charge distributions in an electronic system. However, this technique is quite experimentally challenging, since it requires detector pixels that are small enough to resolve the image of small objects such as molecules.

Another approach has been suggested by Grosser {et al.~}in Reference~\cite{GrosserPRA17}, who have shown that the Compton-scattering cross section, in the impulse approximation, depends solely on the electron momentum distribution. Thus, time-resolved Compton scattering can be used to obtain momentum-space images of the sample to be probed. Kowalewski  et al.~have suggested time-resolved X-ray diffraction as a probe of molecules in the gas phase undergoing nonadiabatic avoided-crossing dynamics involving strongly coupled electrons and nuclei \cite{KowalewskiStrDyn17}. They have shown that it provides signatures of a created electronic coherence on top of dominant ground- and excited-state wavepacket motions. A photon-coincidence measurement based on the time- and wavevector-resolved detection of photons generated by the scattering of multiple X-ray pulses with variable delays has been proposed by Biggs  et al.~in Reference~\cite{BiggsJPhB14}. They have shown that it directly measures multipoint correlation functions of the charge density through superpositions of valence excitations which are created impulsively by the scattering process.





\section{QED Description of Time-Resolved Resonant X-ray Scattering}
\label{SectionRXS}

RXS is an element specific technique that provides insight into charge, orbital and spin degrees of freedom \cite{FinkRPP13, MatsumuraJPSJ13, DmitrienkoActaCrysA05, LoveseyPhRep05}. Due to the resonant nature of this process, the scattering cross section can be considerably enhanced in comparison to the nonresonant case discussed in the previous Section. In~contrast to NRXS, which probes simultaneously electrons involved in the dynamics and electrons that are essentially stationary \cite{DixitJChPh13}, RXS can directly probe (quasi-)particles involved in the dynamics. Therefore, RXS particularly suits for measurement of heavy elements, where the vast majority of electrons are stationary, and resonant energies approach several keV corresponding to \r Angstrom spatial resolution.

RXS is a two-step process consisting of an absorption and an emission process, which involves an intermediate state in distinction to NRXS. In the first step, the absorption of a photon from an X-ray pulse induces a resonant transition of an electron from a core shell to a valence shell of an atomic system ({e.g.}, a molecule or a crystal) being measured. Thus, after the first step, the electronic system is brought from its initial state $I$ to an intermediate state $J_C$ with an electron hole in a core shell of a scattering atom $C$ and an additional electron in the valence shell. In the second step, an electron from either the same valence shell or some other shell fills the electron hole in the core shell leading to a spontaneous emission of a photon, which is detected. The electronic system is in a final state $F$ afterwards. If elastic scattering has taken place, then the final state $F$ coincides with the initial state $I$, and the energy of the scattered photon is equal to $\omega_\i$. An inelastic scattering event results in the final state $F$ being different from the initial state $I$ and leads to scattering of a photon with the energy $\omega_\i-(E_F-E_I)$. As in the case of NRXS, elastic and inelastic scattering events from a stationary system can be distinguished by the spectroscopy of a scattered photon, and the contribution of elastic scattering events to RXS scattering patterns are dominating. Through elastic scattering, resonant scattering patterns are determined by the electron density of the object being measured, but~are indirectly connected to it via the relation \cite{FinkRPP13}
\begin{align}
\frac{dP^{\text{st}\RXS}(\Q)}{d\Omega} \propto\lf|\sum_C f_C e^{i\Q\cdot \mathbf R_C}\rt|^2,\label{EqstRXS}
\end{align}
 where $f_C$ is a scattering amplitude of an atom $C$ located at a position $\mathbf R_C$. Since RXS involves transitions of electrons from core shells, which are highly localized in comparison to X-ray wavelengths, the~spatial distribution of the X-ray electric field is treated within the dipole approximation for each absorbing atom. A scattering amplitude $f_C$ is proportional to the product of the dipole matrix elements describing the absorption process and the emission process: $f_C \propto \la \Phi_I|\boldsymbol\epsilon_s^*\cdot\mathbf r|\Phi_{J_C}\ra \la\Phi_{J_C}|\boldsymbol\epsilon_\i\cdot\mathbf r| \Phi_I\ra$ .

Now, let us consider the interaction of a resonant X-ray pulse with an atomic system with coherent electron dynamics in the valence shell described by the state $|\Psi(t) \ra = \sum_I C_Ie^{-iE_I(t-t_0)} |\Phi_I\ra$. In the absorption step, an electron from a core shell is excited into the valence shell, whereby it destroys the coherent electron dynamics. The system is brought to an intermediate state $J_C$, and the following~step, emission, is determined by this intermediate state and is not connected to the state $|\Psi(t) \ra$. Thus, as in the case of NRXS, it is, first, very unlikely that the final state after the emission process would coincide with the superposition $|\Psi(t) \ra$ and, second, this situation would be spectroscopically indistinguishable. This again demonstrates that the notion of `elastic scattering' for an interaction of a light pulse with a nonstationary electronic system is unclear. Since the relation  in Equation~(\ref{EqstRXS}) relays on the assumption that elastic scattering provides a dominating contribution to a scattering pattern, it must be reconsidered for time-resolved RXS.

The interaction of an ultrashort resonant X-ray pulse with an electronic system evolving coherently has been described in Reference~\cite{PopovaGorelovaPRB15_1} as follows. Since RXS is a two-step process governed by the Hamiltonian $\hat H_{\text{int}}^{(1)}$ in Equation~(\ref{H_int}), the second-order wave function
\begin{align}
|\Psi_{\{n\}}^{\RXS},t_f\ra=&-\int_{t_0}^{t_f}dt'\,e^{i\hat H_0 t'}\hat H_{\text{int}}^{(1)}\,e^{-i\hat H_0 t'}\label{Psi2_App}\\
&\times\int_{t_0}^{t'}dt''\,e^{i\hat H_0 t''}\hat H_{\text{int}}^{(1)}\,e^{-i\hat H_0 t''}|\{n\}\ra\lf|\sum_IC_I\Phi_I\rt\ra,\nonumber
\end{align}
 must be substituted for $|\Psi_{\{n\}},t_f\ra$ in Equation~(\ref{rho_total}). Thus, according to Equations~(\ref{DSP}) and (\ref{ProbabXS}), the DSP of RXS from an electronic system evolving coherently is
%
%
%
\begin{align}
\frac {d P^{\RXS}(\Q)}{d\Omega} =&\frac{1}{4\pi^2 c^3\omega_\i^2}\int_0^{\infty}d\omega_{\ks}\omega_{\k_\s}W(\omega_{\k_\s})\sum_{F,\s}
\int_{t_0}^{+\infty}dt'_1\int_{t_0}^{+\infty}dt'_2\int_{t_0}^{t'_1}dt_1''\int_{t_0}^{t'_2}dt''_2\label{EqDSP_RXS_Gen}\\
&\times\int d^3r_1\int d^3r_2 
G^{(1)}(\br_1,t_1'',\br_2,t_2'') e^{-i\k_\s\cdot(\br_1-\br_2)} M_F^{\RXS}(t_1',t_1'',\br_1)\bigl[M_F^{\RXS}(t_2',t_2'',\br_2)\bigr]^\dagger\nonumber
\end{align}
 with the function
\begin{align}
M_F^{\RXS}(t',t'',\br)=e^{i(E_F-\omega_{\k_\s}) t'}\sum_{J_C}e^{i(E_{J_C}-i\Gamma_{J_C}/2)(t''-t')} \la\Phi_F|\hat T^\dagger_{\s}(\br) |\Phi_{J_C}\ra\la\Phi_{J_C}|  \hat T_{\i}(\br)|\Psi(t'')\ra,
\end{align}
 where $\hat T_{\s(\i)}(\br) = \hat\psi^\dagger(\mathbf r)(\boldsymbol\epsilon_{\s(\i)}\cdot\mathbf p)\hat \psi(\mathbf r)$ and $\Gamma_{J_C}$ is the decay width of the intermediate state $J_C$. This~expression is general for a resonant X-ray pulse with arbitrary coherence properties, and assumes neither the frozen-density approximation nor the dipole approximation. The relation in Equation~(\ref{EqDSP_RXS_Gen}) has much in common with the general expression for time-resolved NRXS from a coherently evolving electronic system in Equation~(\ref{EqDSP_NRXS_Gen}). Analogously to Equation~(\ref{EqDSP_NRXS_Gen}), the DSP for time-resolved RXS involves the sum over final states $F$ with corresponding transition amplitudes summed incoherently and also depends on the first-order correlation function $G^{(1)}(\br_1,t_1'',\br_2,t_2'')$.

Applying the dipole approximation for each absorbing atom, and assuming a coherent resonant X-ray pulse with the amplitude of the electric field defined in Equation~(\ref{EqGaussianPulse}) and duration that is much shorter than the characteristic time scale of the changes in the electron density, the DSP of RXS can be represented as
\begin{align}
&\frac{dP^\RXS(\Q)}{d\Omega}= \mathrm P_0^{\RXS}
\sum_{C_1,C_2}e^{i\mathbf Q\cdot(\mathbf R_{C_1}-\mathbf R_{C_2})}\sum_F \la \Psi(t_p) |  \hat  G_{F,C_1,C_2}^\RXS |\Psi(t_p)\ra e^{-\Omega_F^2\tau_p^2/(4\ln2)},\label{EqDSP_RXS}
\end{align}
 with the operator
\begin{align}
&\hat  G^\RXS_{F,C_1,C_2} =\sum_{\s=1}^2\int_0^{\infty}d\omega_{\k_\s}\omega_{\k_\s}W(\omega_{\k_\s}) \hat S_{C_2}^{\dagger}|\Phi_F\ra\la\Phi_F|\hat S_{C_1}, \label{EqGF_RXS}\\
&\la\Phi_F|\hat S_{C} = \sum_{J_C}\frac{\Delta\omega_{J_{C}F}\la\Phi_F|\boldsymbol\epsilon_\s^*\cdot\hat{\mathbf r}|\Phi_{J_C}\ra\la\Phi_{J_{C}}| \boldsymbol\epsilon_\i\cdot\hat{\mathbf r}}
{\omega_{\k_\s}-E_{J_C}+E_F+i\Gamma_{J_{C}}/2},\nonumber
\end{align}
 where $\mathrm P_0^{\RXS}= \tau_p^2I_0/(4\ln2c^4)$ and $\hat{\mathbf r}  = \int d^3 r \psi^\dagger(\mathbf r)\,\mathbf r\,\psi(\mathbf r)$. The expression in Equation~(\ref{EqDSP_RXS}) is also very similar to the expression for the DSP of NRXS in Equation~(\ref{EqDSP_NRXS}); but, instead of the space integrals in Equation~(\ref{EqDSP_NRXS}), it involves the sum over scattering atoms $C_1$ and $C_2$, and their positions $\mathbf R_{C_1}$ and $\mathbf R_{C_2}$ due to the dipole approximation. In particular, it has a similar temporal dependence on the electronic state at the time of measurement $\Psi(t_p)$.

Let us compare Equation~(\ref{EqDSP_RXS}) with the expression for the DSP from a stationary system in Equation~(\ref{EqstRXS}), which can be represented as $dP^{\text{st}\RXS}/d\Omega \propto \sum_{C_1,C_2}e^{i\mathbf Q\cdot(\mathbf R_{C_1}-\mathbf R_{C_2})} f_{C_1}f^*_{C_2}$. The DSP according to Equation~(\ref{EqDSP_RXS}) could be represented as $dP^{\RXS}/d\Omega \propto \sum_{C_1,C_2}e^{i\mathbf Q\cdot(\mathbf R_{C_1}-\mathbf R_{C_2})}\sum_F \widetilde f_{C_1F}(t_p)\widetilde f^*_{C_2F}(t_p)$, where the scattering amplitudes $\widetilde f_{CF}(t_p)$ are proportional to $\la\Phi_F|\boldsymbol\epsilon_\s^*\cdot\hat{\mathbf r}|\Phi_{J_C}\ra\la\Phi_{J_{C}}| \boldsymbol\epsilon_\i\cdot\hat{\mathbf r}|\Psi(t_p)\ra$. Obviously, it is not possible to obtain the correct expression for the time-resolved DSP by simply substituting the scattering amplitudes $f_{C}\propto \la\Phi_I|\boldsymbol\epsilon_\s^*\cdot\hat{\mathbf r}|\Phi_{J_C}\ra\la\Phi_{J_{C}}| \boldsymbol\epsilon_\i\cdot\hat{\mathbf r}|\Phi_I\ra$ for their time-dependent analogues $f_{C}(t_p)\propto \la\Psi(t_p)|\boldsymbol\epsilon_\s^*\cdot\hat{\mathbf r}|\Phi_{J_C}\ra\la\Phi_{J_{C}}| \boldsymbol\epsilon_\i\cdot\hat{\mathbf r}|\Psi(t_p)\ra$ in Equation~(\ref{EqstRXS}) as
\begin{align}
\frac{dP^{\RXS}}{d\Omega} \propto \lf|\sum_{C}e^{i\mathbf Q\cdot\mathbf R} f_{C}(t_p)\rt|^2.\label{Eq_DSP_RXSwrong}
\end{align}

 This assumption suggests that  the absorption process first brings the coherently evolving electronic system into an intermediate state $J_C$, and then the emission process brings the system from the intermediate state $J_C$ into the same coherent superposition $\Psi(t_p)$ as before the absorption. Equation~(\ref{EqDSP_RXS}) indeed takes into account that the absorption process bringing the system into the intermediate state $J_C$ destroys electron dynamics, and final states after the emission process are eigenstates of the Hamiltonian $\hat H_{\text{m}}$. Therefore, the information about electron dynamics is contained in the absorption term of $\widetilde f_{CF}(t_p)$, but not in the emission term.

\begin{figure}[H]
\centering
\includegraphics[width=8 cm]{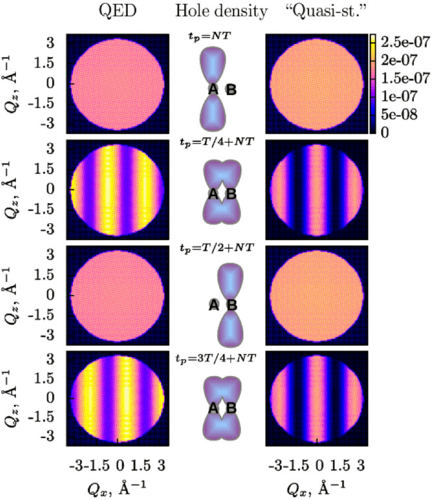}
\caption{DSP from a single Br$_2$ molecule with a coherently oscillating electron hole in the $Q_x-Q_z$ plane at $Q_y=0$ according to Equation~(\ref{EqDSP_RXS}) (left) and Equation~(\ref{Eq_DSP_RXSwrong}) (right) and the corresponding schematic representation of the hole density (middle) at different $t_p$. The Br$_2$ molecule is aligned along the $x$ direction parameters, the interatomic distance is $R_x=2.3$ \r A and the oscillation period $T$ is \mbox{1.7 fs \cite{PottsTrFSoc71}}. Probe pulse is polarized along the $z$ direction and propagates along the $y$-axis with corresponding parameters $\tau_p = 200$ as, $\omega_\i = 13.5$ keV, $I_0 = 10^{18}$ W/cm$^2$. $N$ is a positive integer. The~dependence on polarization is not shown. The ranges are limited by $Q^2_x+Q^2_z\le\omega_\i^2/c^2$. It is assumed that the detector acceptance range $W(\omega_{\ks})$ allows detecting only $1s-4p\pi g$ and $1s-4p\pi u$ scattering events. Figure is reproduced with permission from Reference~\cite{PopovaGorelovaPRB15_1}. Copyright American Physical Society, 2015.
\label{Br2Fig}
}
\end{figure}

The difference between the semiclassical approach to obtain the time-resolved DSP according to Equation~(\ref{Eq_DSP_RXSwrong}) and the correct expression in Equation~(\ref{EqDSP_RXS}) has been illustrated in Reference~\cite{PopovaGorelovaPRB15_1} by the calculation of the DSP from an ionized Br$_2$ molecule with coherent electron dynamics in the valence shell. The two highest occupied molecular orbitals of Br$_2$ are the $\pi g$ and $\pi u$ orbitals of Br $4p$ character. They assumed that a photoionizing pump pulse created an electron hole initially localized at the $4pz$ orbital of one of Br$_2$ atoms (at site A in Figure~\ref{Br2Fig}) at time $t_0=0$, which is actually a coherent superposition of the electron hole being in the $4p\pi g$ orbital and the electron hole being in the $4p\pi u$ orbital. Such electron-hole localization in a molecule by a photoionizing pump pulse is possible as has been demonstrated in Reference~\cite{SansoneNature10}. Then, the electron hole, which was initially localized on one of atoms, starts oscillating coherently between the two atoms of the Br$_2$ molecule moving from site A to site B and back (see~Figure~\ref{Br2Fig}). The electron hole dynamics is probed by an X-ray pulse with the photon energy of 13.5 keV tuned to the $K$ edge of Br.

The DSP from the Br$_2$ molecule has been calculated according to the semiclassical expression in Equation~(\ref{Eq_DSP_RXSwrong}) and the correct expression in Equation~(\ref{EqDSP_RXS}), and compared in Figure~\ref{Br2Fig}. In both cases, the signals are constant when the hole is localized on one of the atoms (at times $0$ and $T/2$), since the scattering channel for the other atom is blocked at these moments, and there is no interference. In contrast to the semiclassical scattering patterns, the scattering patterns according to Equation~(\ref{EqDSP_RXS}) are different from each other at times $T/4$ and $3T/4$, when the electron hole densities are equal, and depend on whether the hole is moving from site A to site B or vice versa. These scattering patterns are not centrosymmetric with respect to $\Q$ as in the case of NRXS (see Section \ref{SectionNRXS}). The~QED patterns at times $T/4$ and $3T/4$ are not just phase-shifted with respect to the semiclassical patterns, but also have different amplitudes. For example, their intensity is nonzero at any $\Q$ in contrast to the semiclassical patterns, which are zero at $Q_xR_x=\pi/2$. This comparison again demonstrates that the semiclassical approach to describe the interaction of a light pulse with a nonstationary electronic system can lead to erroneous time- and momentum-dependence of a time-resolved signal. Correct results demand a thorough analysis within the QED framework.

The analysis of a feasibility to realize the experiment as described above has been performed \mbox{in Reference~\cite{PopovaGorelovaPRB15_1}}. It has been shown that if one would prepare a beam of aligned molecules as shown in References~\cite{HolmegaardPRL09, KupperPRL14} with the density of 10$^{10}$ cm$^{-3}$ and size 0.4 cm, and excite a coherent electron dynamics in 10\% of the molecules, then approximately four molecules per shot would contribute to a scattering pattern assuming the interaction area of 10$^{-8}$ cm$^2$. Thus, summation of signals from roughly 10$^5$~scattering patterns provides a signal of 0.01 photons per pixel, which are enough to reconstruct a structure of a single molecule \cite{FungNature08}. This means that such an experiment is in principle feasible with the forthcoming European X-ray free electron laser facility, which will provide 27,000 X-ray pulses per second \cite{BartyARPC13} and allow collecting the necessary number of scattering patterns within hours.

The scattering patterns in Figure~\ref{Br2Fig} depend on the direction in which the electron hole is moving. This means that the scattering patterns encode the interatomic electron current and it must be possible to reconstruct it. It has been shown in References~\cite{PopovaGorelovaPRB15_1, PopovaGorelovaPRB15_2} that this is a general property of scattering patterns obtained by means of ultrafast RXS and developed the following method to reconstruct the interatomic electron current. We applied the analogous analysis to that employed by this method for the case of NRXS in the previous Section \ref{SectionNRXS} and obtained very similar results.

Equation~(\ref{EqDSP_RXS}) can be represented as
\begin{align}
\frac{dP^{\RXS}(\Q)}{d\Omega} =&\sum_{C_1,C_2} \cos[\Q\cdot(\mathbf R_{C_1}-\mathbf R_{C_2})] \mathcal R^{\RXS}_{C_1,C_2}(t_p)\label{EqDSP_RXS_JR}-\sum_{C_1,C_2} \sin[\Q\cdot(\mathbf R_{C_1}-\mathbf R_{C_2})] \mathcal J_{C_1,C_2}^{\NRXS}(t_p),\nonumber
\end{align}
 where the first term is centrosymmetric with respect to $\Q$ and is determined by the fuction
\begin{align}
\mathcal R^{\RXS}_{C_1,C_2}(t_p) =& \mathrm P_0^{\RXS}  \Re\lf[\la\Psi(t_p)|\sum_F\hat G_{F,C_1,C_2}^{\RXS}|\Psi(t_p)\ra \rt]e^{-\Omega_F^2\tau_p^2/(4\ln2)},
\end{align}
  which depends on the real parts of the density matrix elements and, therefore, evolves in time similarly to the electron density [{cf.}~Equation~(\ref{EqElDensity})] and charge distributions.
The second term is an odd function with respect to $\Q$ and is determined by the function
\begin{align}
\mathcal J^{\RXS}_{C_1,C_2}(t_p) =& \mathrm P_0^{\RXS}  \Im\lf[\la\Psi(t_p)|\sum_F\hat G_{F,C_1,C_2}^{\RXS}|\Psi(t_p)\ra \rt]e^{-\Omega_F^2\tau_p^2/(4\ln2)}
\end{align}
 connected to the imaginary parts of the density matrix elements and correlates with the probability current density $\mathbf j(\mathbf r, t_p)$ [{cf.}~Equation~(\ref{ProbCurrDen})]. $\mathbf j(\mathbf r, t_p)$ can be decomposed into intra-atomic and inter-atomic contributions as
\begin{align}
&\mathbf j(\mathbf r, t_p) = \sum_{C} \mathbf j_{C}^{\text{intra}}(\mathbf r, t_p) + \sum_{C_1,C_2\neq C_1} \mathbf j_{C_1C_2}^{\text{inter}}(\mathbf r, t_p),
\end{align}
 where the sums are over all atoms in the system. The volume-integrated probability current between scattering atoms $C_1$ and $C_2$ is given by
\begin{align}
j_{C_1C_2}(t_p) &= \int d^3 r \, \mathbf j_{C_1C_2}^{\text{inter}}(\mathbf r, t_p)\cdot\mathbf n_{C_1C_2} \label{InterAtCurr}\\
 &= \Im\la\Psi(t_p)|\hat {\mathcal G}_{C_1C_2}|\Psi(t_p)\ra\nonumber,
\end{align}
 where $\mathbf n_{C_1C_2}$ is the unit vector pointing from site $C_1$ to site $C_2$, and $\hat {\mathcal G}_{C_1C_2} =  \int d^3 r \hat\xi_{C_2}^\dagger(\mathbf r)(\boldsymbol\nabla\cdot\mathbf n_{C_1C_2})\hat\xi_{C_1}(\mathbf r)$ with the operator $\hat\xi_{C}$ annihilating a particle at site $\mathbf R_C$. It has been shown in Reference~\cite{PopovaGorelovaPRB15_1}  that the time evolution of the interatomic electron current $j_{C_1C_2}(t_p)$ is substantially reproduced by the function $\mathcal J^{\RXS}_{C_1,C_2}(t_p)$ and can be reconstructed from the Fourier transform of the DSP as shown below.


Performing the Fourier transform of the DSP in Equation~(\ref{EqDSP_RXS_JR}), $\mathcal F^{\RXS}(\br,t_p ) = 1/(2\pi)^3\int d^3Q e^{-i\Q\cdot\br}dP^{\RXS}(\Q)/d\Omega$, its centrosymmetric and noncetrosymmetric contributions can be decomposed. The real part of the Fourier transform is connected to the functions $\mathcal R_{C_1,C_2}^{\RXS}$ via the~relation
\begin{align}
\Re\lf[ \mathcal F^{\RXS}(\br,t_p)\rt] =  \sum_{C_1,C_2}\mathcal R^{\RXS}_{C_1,C_2}(t_p)\delta\lf[\br-(\mathbf R_{C_1}-\mathbf R_{C_2})\rt],
\end{align}
 and its imaginary part is connected to $\mathcal J_{C_1,C_2}^{\RXS}$ via
\begin{align}
\Im\lf[ \mathcal F^{\RXS}(\br,t_p)\rt] =  \sum_{C_1,C_2}\mathcal J^{\RXS}_{C_1,C_2}(t_p)\delta\lf[\br-(\mathbf R_{C_1}-\mathbf R_{C_2})\rt].\label{EqImFourierRXS}
\end{align}

 Thus, the Fourier transform of a time-resolved resonant scattering pattern is a sum of delta peaks at the interatomic distances  $\mathbf R_{C_1}-\mathbf R_{C_2}$ weighted by complex amplitudes, which depend on the time of measurement $t_p$. In particular, the imaginary part of the amplitude of a peak at $\mathbf R_{C_1}-\mathbf R_{C_2}$ is given by $\mathcal J^{\RXS}_{C_1,C_2}(t_p)$ and provides the electron current between atoms $C_1$ and $C_2$. If there are several atomic pairs at the same interatomic distance, when the imaginary part corresponds to the sum of the interatomic currents between the corresponding pairs.

\begin{figure}[H]
\centering
\includegraphics[width=6 cm]{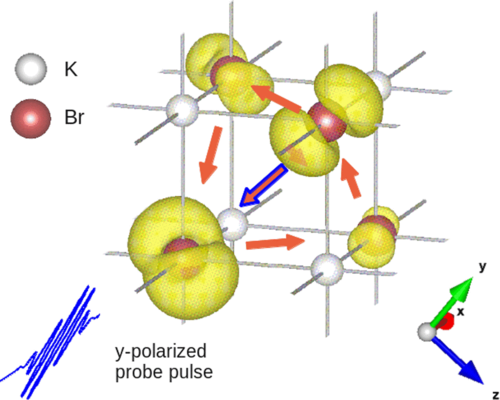}
\caption{Snapshot of the electron-hole density at time $t_p$ in a fragment of (KBr)$_{108}$ cluster visualized using VESTA software \cite{MommaJAC11}. The orange arrows represent the electron-hole currents $j_{C_1C_2}(t_p)$ between the Br atoms. The blue-framed arrow represents a current parallel to the probe-pulse polarization. Figure is reproduced with permission from Reference~\cite{PopovaGorelovaPRB15_2}. Copyright American Physical Society, 2015.
\label{KBrFig}
}
\end{figure}

The procedure to reconstruct interatomic electron currents in crystals by means of ultrafast RXS has been illustrated by the calculation of time-resolved scattering patterns from KBr and Ge crystals with coherent electron dynamics in the valence bands by D.~Popova-Gorelova and R.~Santra in Reference~\cite{PopovaGorelovaPRB15_2}. Here, we will review their calculation of coherent electron dynamics in KBr. It is an ionic crystal, where the $4s$ electrons of K atoms are transferred to Br atoms \cite{WertheimPRB95}. The $p$-character electrons centered on the Br atoms form the outermost valence band of KBr. It was assumed that a photoionizing pump pulse induced coherent electron dynamics in this band by creating $p$-type electron holes centered on Br atoms at time $t_0$. A probe pulse propagating along the $x$ direction of the duration of 200 as and the photon energy of $\approx13$ keV tuned to the $K$ edge of Br polarized along the $y$-axis parallel to one of the vectors connecting two Br atoms (see Figure~\ref{KBrFig}) was assumed.

Each coherently evolving electron hole is delocalized and distributed over many Br atoms in some region (see Figure~\ref{KBrFig}). It was assumed that the concentration of the electron holes is sufficiently low to consider these regions isolated and the holes noninteracting. A region, where a single electron hole is distributed, was simulated by a (KBr)$_{108}$ cluster. Since the delocalized electron hole is coherently evolving, the charge is redistributing and flowing from one atom to another resulting in nonzero interatomic electron current.

\begin{figure}[H]
\centering
\includegraphics[width=6 cm]{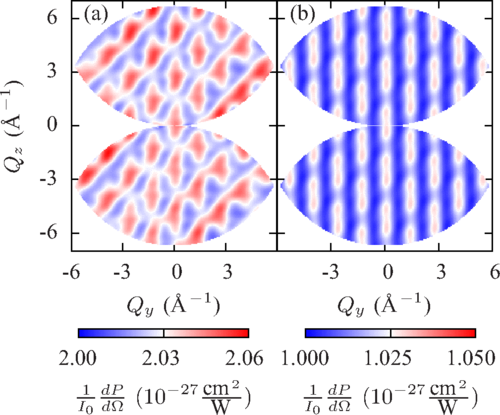}
\caption{Scattering patterns at the probe-pulse intensity $I_0$ from the ionized KBr cluster obtained by a $y$-polarized X-ray pulse. (\textbf{a}) No polarization filter is applied in the measurement of a scattered photon; (\textbf{b}) A polarization filter transmitting $y$-polarized scattered photons is applied and the dependence on scattered polarizations is eliminated. The spectral window function $W(\omega_\ks)$ is centered at $\omega_\i$ and suppresses all photons emitted by electrons lying deeper than the outermost valence band of KBr. Figure is reproduce with permission from Reference~\cite{PopovaGorelovaPRB15_2}. Copyright American Physical Society, 2015.
\label{KBrPattFig}
}
\end{figure}

\begin{figure}[H]
\centering
\includegraphics[width=6 cm]{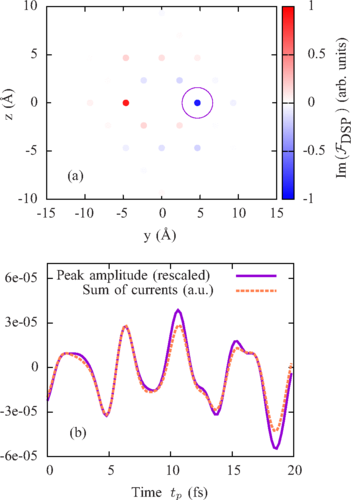}
\caption{(\textbf{a}) The imaginary part of the Fourier transform from $\Q$ space to real space of the scattering pattern in Figure~\ref{KBrPattFig}b; (\textbf{b}) Solid violet line: time evolution of the amplitude of the circled peak in panel~(\textbf{a}). Orange dashed line: the sum of the currents between pairs of atoms connected by the vector $(0,R_{\text{Br-Br}},0)$, where $R_{\text{Br-Br}}$ is the interatomic distance between nearest-neighbour Br atoms in KBr. Figure is reproduced with permission from Reference~\cite{PopovaGorelovaPRB15_2}. Copyright American Physical Society, 2015.
\label{InterAtCurrFig}
}
\end{figure}

A scattering pattern at a time $t_p$ is shown in Figure~\ref{KBrPattFig}. Since scattered polarizations are different at each $\ks$ vector, an additional $\Q$-dependence appears in the DSP and leads to a nonperiodic scattering pattern in Figure~\ref{KBrPattFig}a. This dependence is eliminated in the scattering pattern in Figure~\ref{KBrPattFig}b, which is determined solely by the trigonometric functions in Equation~(\ref{EqDSP_RXS_JR}). Figure~\ref{InterAtCurrFig}a shows the imaginary part of the Fourier transform from $\Q$ space to real space of the scattering pattern in Figure~\ref{KBrPattFig}b extrapolated to a region of infinite $Q_y$ and $Q_z$. In accordance with Equation~(\ref{EqImFourierRXS}), it consists of delta peaks at points corresponding to vectors connecting pairs of Br atoms (see Figure~\ref{KBrFig}). The amplitudes of peaks at points $(R_y,R_z)$ are opposite to the amplitudes of peaks at $(-R_y,-R_z)$.

Performing an analysis of the electronic structure of KBr, the authors found a way to resolve interatomic currents between Br atoms in a direction parallel to the probe-pulse polarization. They~suggested to measure scattering patterns at different time delays $t_p-t_0$, perform the Fourier transform of these scattering patterns from $\Q$ space to real space and follow the amplitude of the peak outlined in the circle in Figure~\ref{InterAtCurrFig}a in the imaginary part of the Fourier transform. The evolution of the amplitude of this peak depending on the time of measurement $t_p$ is shown in Figure~\ref{InterAtCurrFig}b with the solid violet line. It precisely reproduces the time evolution of the computed sum of the currents between all pairs of nearest-neighbor Br atoms connected by the vector $(0,R_{\text{Br-Br}},0)$ shown in the orange dashed line in Figure~\ref{InterAtCurrFig}b, where $R_{\text{Br-Br}}$ is the distance between the nearest-neighbor Br atoms. Thus, it has been shown in Reference~\cite{PopovaGorelovaPRB15_2} that if a proper polarization of the incoming probe pulse has been chosen for measurement of scattering patterns, the time evolution of the amplitude of a certain delta peak in their Fourier transform follows the time evolution of the sum of the interatomic currents between atoms connected by the vector corresponding to this peak. The required polarization of the probe pulse can be determined by an analysis of the electron structure of the crystal.

To sum up, time-resolved RXS is a robust technique that is in many ways advantageous for probing electron dynamics in molecules and crystals. On top of the usual structural information about an atomic system, it can provide interatomic electron current in it. Feasibility analyses of the experiments to measure electron dynamics in Br$_2$ \cite{PopovaGorelovaPRB15_1}, and KBr and Ge crystals \cite{PopovaGorelovaPRB15_2} by means of ultrafast RXS showed that they are achievable at X-ray free electron laser facilities. However, the main obstacle of such experiments is the need to produce hard X-ray pulses of subfemtosecond duration for an interatomic spatial resolution and sufficient temporal resolution. Although generation of such pulses has been recently achieved \cite{HuangPRL17}, these constraints on the probe-pulse parameters are still quite demanding. These restrictions can be partially overcome with TRARPES technique \cite{Popova-GorelovaPRA16}, which we review in the next Section.

\section{QED Description of Time-Resolved Photoelectron Probability}
\label{SectionPE}

TRARPES,  i.e., time- and energy-resolved molecular-frame photoelectron angular distributions, has been proposed for imaging coherent electron dynamics in molecules \cite{MignoletPRA12, KusJPCA13, PerveauxJPCA14, Popova-GorelovaPRA16}. An advantage of this technique over X-ray scattering is that it allows achieving \r Angstrom spatial resolution with light pulses of a much lower photon energy. An analysis of time- and angle-resolved photoelectron distributions obtained by ultrashort XUV probe pulses inducing single-photon ionization of a coherently evolving electron system within the QED framework has been performed in Reference~\cite{Popova-GorelovaPRA16}. This study provided a correct way to take into account the consequence of the broad probe-pulse bandwidth on a signal, which has not been taken into account in earlier studies of time-resolved photoelectron probability~\cite{MignoletPRA12}, but is critical for a correct interpretation of time-resolved photoelectron spectra. This work is reviewed in this Section.

In order to describe the absorption process, the total density matrix must be evaluated within the first-order time-dependent perturbation theory using the Hamiltonian $\hat H_{\text{int}}^{(1)}$. Thus, in order to obtain the photoelectron probability, the first-order wave function
\begin{align}
|\Psi_{\{n\}}^{\PE},t_f\ra=&-i\int_{t_0}^{t_f}dt\,e^{i\hat H_0 t}\hat H_{\text{int}}^{(1)}\,e^{-i\hat H_0 t}|\{n\}\ra\lf|\sum_I C_I\Phi_I\rt\ra\label{Psi1}
\end{align}
 must be substituted for $|\Psi_{\{n\}},t_f\ra$ in Equation~(\ref{rho_total}). Thus, according to Equation~(\ref{ProbabPE}), the photoelectron probability of a coherently evolving electronic system consisting of $N_{\el}$ electrons is
\begin{align}
P^\PE(\Q_\el)  =\frac{1}{\omega_\i^2} \sum_{F,\sigma}&\int_{t_0}^{t_f} dt_1\int_{t_0}^{t_f} dt_2\int d^3r_1\int d^3 r_2\,\phi_{\sigma e}^\dagger(\Q_\el,\br_1)\phi_{\sigma e}(\Q_\el,\br_2)\label{EqPE_gen}\\
&\times G^{(1)}(\br_1,t_1,\br_2,t_2)  M_F^{\PE}(t_1,\br_1)\bigl[M_F^{\PE}(t_2,\br_2)\bigr]^\dagger\nonumber
\end{align}
 with the function
\begin{align}
M_F^{\PE}(t,\br)=&e^{i (E_F^{N_\el-1}+\varepsilon_e) t}\la \Phi_F^{N_\el-1} |(\boldsymbol\epsilon_{\i}\cdot\mathbf p)\hat \psi(\mathbf r)|\Psi(t)\ra.
\end{align}

Here, $\phi_{\sigma e}(\Q_\el,\br)$ is the wave function of a photoelectron with spin $\sigma$ and momentum $\Q_\el$, \mbox{$\varepsilon_e = |\Q_\el|^2/2$} is the photoelectron energy, $|\Phi_F^{N_\el-1}\ra$ is a final state of the electronic system with $N_\el-1$ electrons and energy $E_F^{N_{\el}-1}$, which by assumption does not interact with the emitted photoelectron. Equation~(\ref{EqPE_gen}) does not include any further assumptions concerning probe-pulse parameters, such as coherence properties or duration.

Applying the dipole and the frozen-density approximation, assuming a coherent XUV pulse with the amplitude of the electric field defined in Equation~(\ref{EqGaussianPulse}), and applying the plane-wave approximation to the photoelectron wave function, the photoelectron probability can be represented as
\begin{align}
P^\PE(\Q_\el) =&\mathrm P_0^{\PE}|\boldsymbol\epsilon_\i\cdot\Q_{\el}|^2\sum_{F,\sigma} \int d^3r_1\int d^3 r_2 
e^{i\Q_\el\cdot(\br_2-\br_1)}\la\Psi(t_p)|\hat G_F^\PE(\br_1,\br_2)|\Psi(t_p)\ra e^{-(\Omega_{F}^\el-\varepsilon_e)^2\tau_p^2/(4\ln2)}\label{EqPE}
\end{align}
 with the operator
\begin{align}
\hat G_F^\PE(\br_1,\br_2)= \hat\psi^\dagger(\br_2)|\Phi_F^{N_{\el}}\ra\la\Phi_F^{N_{\el}}|\hat\psi^\dagger(\br_1), \label{EqGF_PE}
\end{align}
 $\mathrm P_0^{\PE}= I_0\tau_p^2/(8\pi\ln2\omega_\i^2c)$ and $\Omega_F^\el = \omega_\i-E_F^{N_{\el}-1}+\la E \ra$.

The general expression Equation~(\ref{EqPE_gen}) for the time- and angle-resolved photoelectron probability has much in common with the general expressions for the DSPs of time-resolved NRXS [Equation~(\ref{EqDSP_NRXS_Gen})] and time-resolved RXS [Equation~(\ref{EqDSP_RXS_Gen})], as well as the expression in Equation~(\ref{EqPE}) is analogous to the corresponding expressions in Equations~(\ref{EqDSP_NRXS}) and (\ref{EqDSP_RXS}). The general \mbox{Equations~(\ref{EqDSP_NRXS_Gen}), (\ref{EqDSP_RXS_Gen}) and (\ref{EqPE_gen})} depend on the first-order radiation field correlation function $G^{(1)}(\br_1,t_1,\br_2,t_2)$ multiplied by a function dependent on $\Psi(t_1)$ and its conjugate dependent on $\Psi(t_2)$. Equations~(\ref{EqDSP_NRXS}), (\ref{EqDSP_RXS}) and (\ref{EqPE}) describe analogous dependence of the corresponding time-resolved signals on the evolution of the electronic state $\Psi(t_p)$ and the spectral density of an ultrashort probe pulse. All these relations involve an incoherent sum over final states. The only prominent difference of the relations for the photoelectron probability from the relations for X-ray scattering is that the former comprise the photoelectron wave function $\phi_{\sigma e}(\Q_\el,\br)$ instead of the wave function of a scattered photon $e^{i\ks\cdot \br}$.

In addition, analogously to time-resolved NRXS and time-resolved RXS, there are two contributions to the photoelectron probability in Equation~(\ref{EqPE}), which can be represented as
\begin{align}
P^\PE(\Q_\el) =&|\boldsymbol\epsilon_\i\cdot\Q_{\el}|^2 \int d^3r_1\int d^3 r_2 
\cos[\Q_\el\cdot(\br_2-\br_1)]\mathcal R^\PE(t_p,\varepsilon_e,\br_1,\br_2)\label{EqPE_RJ}\\
&-|\boldsymbol\epsilon_\i\cdot\Q_{\el}|^2 \int d^3r_1\int d^3 r_2 
\sin[\Q_\el\cdot(\br_2-\br_1)]\mathcal J^\PE(t_p,\varepsilon_e,\br_1,\br_2).\nonumber
\end{align}

The first contribution is centrosymmetric with respect to $\Q_\el$ and is determined by the function
\begin{align}
\mathcal R^\PE(t_p,\varepsilon_e,\br_1,\br_2) = \mathrm P_0^{\PE}\Re\lf[\la\Psi(t_p)|\sum_F\hat G_F^\PE(\br_1,\br_2)|\Psi(t_p)\ra \rt] e^{-(\Omega_{F}^\el-\varepsilon_e)^2\tau_p^2/(4\ln2)},\label{Eq_R_PE}
\end{align}
 which is connected to the real parts of electron density matrix elements. The second contribution is an odd function with respect to $\Q_\el$ and is determined by
\begin{align}
\mathcal J^\PE(t_p,\varepsilon_e,\br_1,\br_2) = \mathrm P_0^{\PE}\Im\lf[\la\Psi(t_p)|\sum_F\hat G_F^\PE(\br_1,\br_2)|\Psi(t_p)\ra \rt] e^{-(\Omega_{F}^\el-\varepsilon_e)^2\tau_p^2/(4\ln2)},\label{Eq_J_PE}
\end{align}
 which is dependent on the imaginary parts of electron density matrix elements. The time- and angle-resolved photoelectron probability is a sum of an even function with respect to $\Q_\el$ determined by $\mathcal R^\PE(t_p,\br_1,\br_2)$ and an odd function with respect to $\Q_\el$ determined by $\mathcal J^\PE(t_p,\br_1,\br_2)$ not only in the case of the photoelectron wave function being a plane wave, but in the more general case of $\phi_{\sigma e}(\Q_\el,\br)$ being a Hermitian function with respect to both $\Q_\el$ and $\br$.

By an analogy to the concept of a chemical shift in stationary photoelectron spectroscopy, one~may assume that electron dynamics can be measured by means of time-dependent chemical shifts,  i.e., that~photoelectron peaks would shift following the time-dependent electron density in a time-resolved measurement. However, according to Equation~(\ref{EqPE}), time-dependent photoelectron spectra at each emission angle consist of a series of photoelectron peaks centered at energies $\Omega_F$ corresponding to a transition to a final state $F$, whereby the position of the peaks are time independent. Therefore, quite counter-intuitively, electron dynamics cannot be followed by means of time-dependent chemical shifts. The quantity that depends on the electronic state at the time of measurement is the strength of the photoelectron peaks.

\begin{figure}[H]
\centering
\includegraphics[width=6 cm]{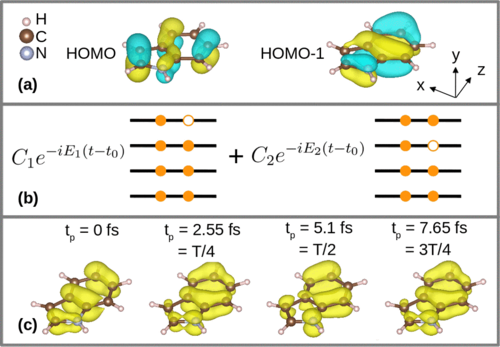}
\caption{(\textbf{a}) Isosurfaces of the amplitudes of the HOMO and HOMO-1 orbitals of indole calculated with the software package the MOLCAS package \cite{KarlstromPSSD03}; (\textbf{b}) Schematic representation of the electronic state of indole after the pump pulse arriving at time $t_0$, which is the superposition of two states with an electron hole in HOMO and HOMO-1 orbitals; (\textbf{c}) Electron hole density at time 0, $T/4$, $T/2$, and $3T/4$. The orbitals and electron hole densities are visualized using the VESTA software \cite{MommaJAC11}. Figure is reproduced with permission from Reference~\cite{Popova-GorelovaPRA16}. Copyright American Physical Society, 2016.
\label{IndoleFig}
}
\end{figure}

Equation~(\ref{EqPE}) was applied for the calculation of time- and angle-resolved photoelectron spectra of an indole molecule with coherent electron dynamics in valence orbitals. It was assumed that a broadband photoionizing pump pulse launched coherent electron dynamics in indole by creating an electron hole in a superposition of the HOMO (highest occupied molecular orbital) and HOMO-1 orbitals at time $t_0$ (see Figure~\ref{IndoleFig}). Then, the electronic state of the indole molecular cation after the interaction with the pump pulse evolves in time as $|\Psi(t)\ra = C_1e^{-iE_1(t-t_0)}|\Phi_H^{\text{ion}}\ra+C_2e^{-iE_2(t-t_0)}|\Phi_{H-1}^{\text{ion}}\ra$, where $|\Phi_H^{\text{ion}}\ra$ is an electronic state with an electron hole in the HOMO, $|\Phi_{H-1}^{\text{ion}}\ra$ is an electronic state with an electron hole in the HOMO-1, and $C_1$ and $C_2$ are time-independent complex coefficients, which are determined by the pump process. After the pump process, the electron density starts oscillating with the period $T=2\pi/(E_2-E_1)\approx 10.2$ fs. Since there are two electronic states involved in the dynamics, there are two time points during the density oscillation period, at which the time-dependent electron densities are identical. The zero time point was adjusted such that the time-dependent electron densities coincide at times $t=T/4$ and $t=3T/4$.

It was suggested to probe the electron dynamics by an XUV pulse, which creates a second electron hole in the indole molecule by a single-photon ionization at time $t_p$, and analyze angular distributions of photoemission probabilities at a fixed photoelectron energy depending on the time delay $t_p-t_0$. A $y$-polarized probe pulse of 1 fs duration, 100 eV photon energy, and 10$^{12}$ W/cm$^2$ intensity was considered. An application of the XUV probe pulse provided photoelectron angular distributions at photoelectron energies of $\sim$ 80 eV--90 eV, which allowed for \r Angstrom spatial resolution. In~addition, an~analysis of such distributions was simplified, since the orthogonalization correction to the plane-wave approximation was suppressed at these energies in the case of the indole molecule.

\begin{figure}[H]
\centering
\includegraphics[width=15 cm]{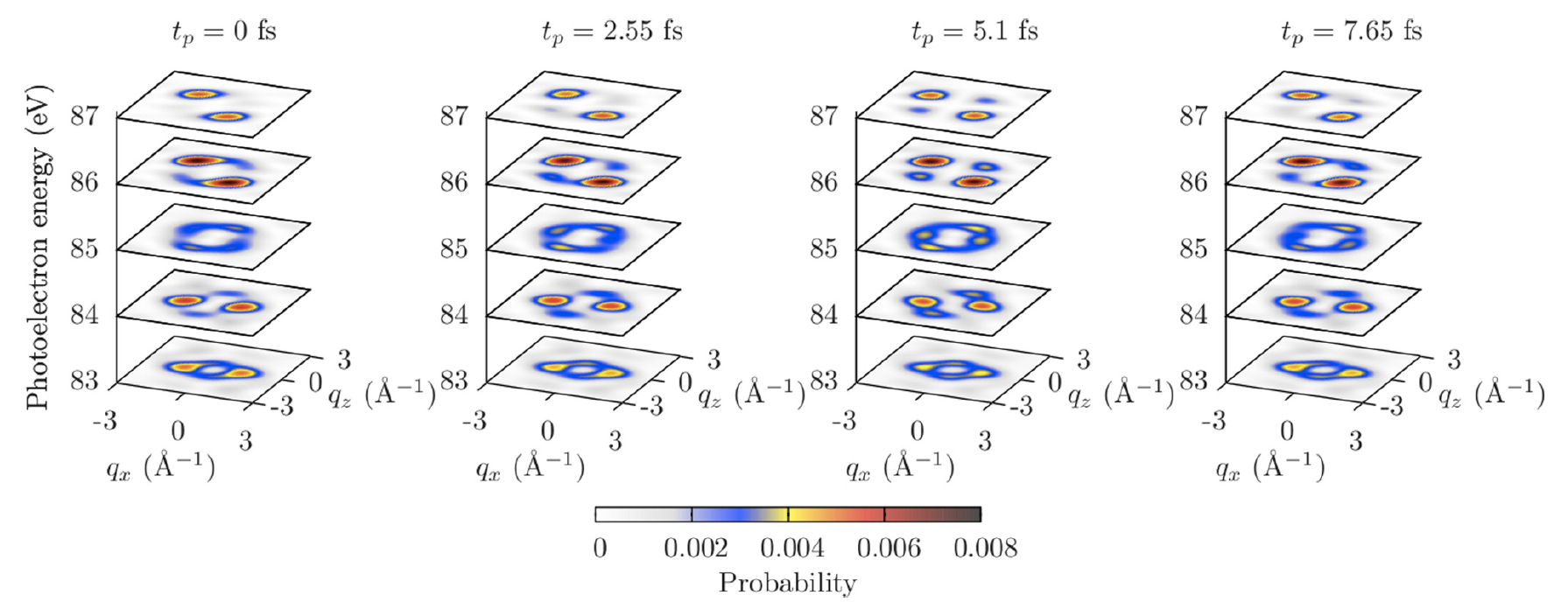}
\caption{Angle-resolved photoelectron spectra generated by the probe pulse arriving at time $t_p$. Each~$q_x-q_z$ plane at the corresponding photoelectron energy $\varepsilon_e$ is a projection of the semisphere with radius $|\mathbf q| = \sqrt 2 \varepsilon_e$ at $q_y>0$, where the color of the $\mathbf q$ point on the semisphere corresponds to the probability of detecting an electron with momentum $\mathbf q$. Figure is reproduced with permission from Reference~\cite{Popova-GorelovaPRA16}. Copyright American Physical Society, 2016.
\label{SpectraFig}
}
\end{figure}

Figure~\ref{SpectraFig} shows calculated angular distributions of photoemission probabilities at different photoelectron energies depending on time $t_p$. The distributions at energies $\varepsilon_e = 85$ eV and 86~eV show quite strong dependence on the electron dynamics. Analogously to the nonresonant and the resonant scattering patterns presented in Sections \ref{SectionNRXS} and \ref{SectionRXS}, correspondingly, the angular distributions do not coincide at times $t_p=T/4=2.55$ fs and $t_p=3T/4=5.1$ fs, when the electron densities are equal, but~the phases of the electronic wave packet are different (see Figure~\ref{IndoleFig}).


It was suggested to analyze the angular distributions at a fixed photoelectron energy $\varepsilon_e$ by performing the Fourier transform according to the following equation
\begin{align}
\mathcal F^\PE(\mathbf r,\varepsilon_e,t_p) &=\int d^3 Q_{\el} \frac{P(\Q_\el,t_p)}{|\boldsymbol\epsilon_\i\cdot\Q_\el|^2}e^{i \Q_\el \cdot\mathbf r}\delta(|\Q_\el|-Q_0)\label{F_2D}
\end{align}
 where $\delta(|\Q_\el|-Q_0)$ is the Dirac delta function. This equation describes the Fourier transform of the spherical surfaces in $\Q_\el$ space of a fixed radius $Q_0 = \sqrt{2\varepsilon_e}$, the projections of which are shown in Figure~\ref{SpectraFig}. It was found that the Fourier transform is connected to the time-dependent Dyson orbitals $\phi_F^D(\br,t_p) = \la\Phi_F^{N_\el-1}|\hat\psi(\br)|\Psi(t_p)\ra$ via
\begin{align}
\mathcal F^\PE(\mathbf r,\varepsilon_e,t_p) &= 8\pi\varepsilon_e\mathrm P_0^\PE\sum_{F,\sigma}e^{-(\Omega_F-\varepsilon_e)^2\tau_p^2/(4\ln2)}\lf(A[\phi_F^D(t_p)] *s\rt)(\br),\label{F_2D_Dyson}
\end{align}
 where $A[\phi_{F}^D(t_p)] = \int d^3 r'{\phi_{F}^D}^\dagger(\mathbf r'-\mathbf r,t_p)\phi_{F}^D(\mathbf r',t_p) $ is the autocorrelation function of $\phi_{F}^D(\mathbf r,t_p)$, $(A[\phi_F^D(t_p)]*s)(\mathbf r)=\int d^3r'A(\mathbf r')s(\mathbf r-\mathbf r')$ denotes the convolution of the autocorrelation function $A[\phi_F^D(t_p)]$ with the function $s(\mathbf r)=\operatorname{sinc}(Q_0|\mathbf r|)$. Thus, the Fourier transform of the time- and angle-resolved photoelectron probability at a fixed photoelectron energy $\varepsilon_e$ is determined by a linear combination of autocorrelation functions of Dyson orbitals $A[\phi_{F}^D(t_p)]$ with the coefficients given by the exponential factor $e^{-(\Omega_F-\varepsilon_e)^2\tau_p^2/(4\ln2)}$. 

Alternatively to the representation of $\mathcal F^\PE(\mathbf r,\varepsilon_e,t_p) $ via the Dyson orbitals in Equation~(\ref{F_2D_Dyson}) provided in Reference~\cite{Popova-GorelovaPRA16}, the Fourier transform can be expressed via the functions $\mathcal R^\PE$ and $\mathcal J^\PE$ defined in Equations~(\ref{Eq_R_PE}) and (\ref{Eq_J_PE}) as
\begin{align}
\Re\lf[\mathcal F^\PE(\mathbf r,\varepsilon_e,t_p)\rt] &= 8\pi\varepsilon_e \int d^3 r' \mathcal R^\PE(t_p,\varepsilon_e,\br',\br'-\br),\\
\Im\lf[\mathcal F^\PE(\mathbf r,\varepsilon_e,t_p)\rt] &= 8\pi\varepsilon_e \int d^3 r' \mathcal J^\PE(t_p,\varepsilon_e,\br',\br'-\br).
\end{align}

Thus, analogously to the results of the Fourier analysis of scattering patterns, the real part of the Fourier transform of the photoelectron angular distributions is determined by the real part of electron density matrix elements, and correlates with the electron density and related quantities such as charge distributions at the time of the probe-pulse arrival. The imaginary part of the Fourier transform is determined by the imaginary part of electron density matrix elements and correlates with the instantaneous electron current. 


\section{Applicability of the Frozen-Density Approximation}
\label{SectionFrozenDensity}

Let us now consider the treatment of electron dynamics during the interaction of an electronic system with the probe pulse. In all examples discussed in the previous Sections, duration of a probe pulse was chosen such short that changes in the electron density during the probe were negligible. As a result, the frozen-density approximation, which neglects electron dynamics during the interaction with the probe pulse, could be applied for the QED derivation of Equation~(\ref{EqDSP_NRXS}) for time-resolved NRXS, of Equation~(\ref{EqDSP_RXS}) for time-resolved RXS, and~of Equation~(\ref{EqPE}) for the time-resolved photoelectron probability. Time-resolved signals according to these expressions are connected to the electronic state $\Psi(t_p)$ at the time of measurement.  


In order to check the applicability of the frozen-density approximation, a situation when the probe-pulse duration is comparable to the characteristic time scale of electron dynamics was considered in Reference~\cite{Popova-GorelovaPRA16}. It was assumed that the pump pulse launched coherent electron dynamics by creating an electron hole in a superposition of HOMO, HOMO-1, and HOMO-2 of indole, which was measured by the probe pulse of the same parameters as in the previous section. The shortest beating period of the time-dependent electron density in this case was 1.5 fs, which was still larger than the probe-pulse duration of 1 fs. 

In order to accurately describe the situation, when the probe pulse duration is comparable with or longer than the characteristic time scale of electron dynamics, one has to take into account the evolution of the electronic system during the action of the probe pulse. The corresponding relation for TRARPES can be derived from the general expression in Equation~(\ref{EqPE_gen}), resulting in

\begin{align}
P^\PE(\Q_\el) =&\mathrm P_0^{\PE}|\boldsymbol\epsilon_\i\cdot\Q_{\el}|^2\sum_{F,\sigma} \int d^3r_1\int d^3 r_2 
e^{i\Q_\el\cdot(\br_2-\br_1)}\label{EqPE_nfrzd}\\
&\qquad\times \sum_{I,K}\la\Psi_K(t_p)|\hat G_F^\PE(\br_1,\br_2)|\Psi_I(t_p)\ra e^{-(\Omega_{FK}^\el-\varepsilon_e)^2\tau_p^2/(8\ln2)} e^{-(\Omega_{FI}^\el-\varepsilon_e)^2\tau_p^2/(8\ln2)},\nonumber
\end{align}
 where $\Psi_{I(K)}(t_p) = \Phi_{I(K)}e^{-iE_{I(K)}(t_p-t_0)}$ and $\Omega^\el_{FI(FK)}=\omega_\i-E_F^{N_{\el}-1}+E_{I(K)}$. Please notice that this relation does not depend on the electronic state at the time of measurement $\Psi(t_p)=\sum_IC_I\Phi_{I}e^{-iE_{I}(t_p-t_0)}$, since the corresponding sum cannot be singled out due to the factors $e^{-(\Omega_{FI(FK)}^\el-\varepsilon_e)^2\tau_p^2/(8\ln2)}$.
 
\begin{figure}[H]
\centering
\includegraphics[width=5 cm]{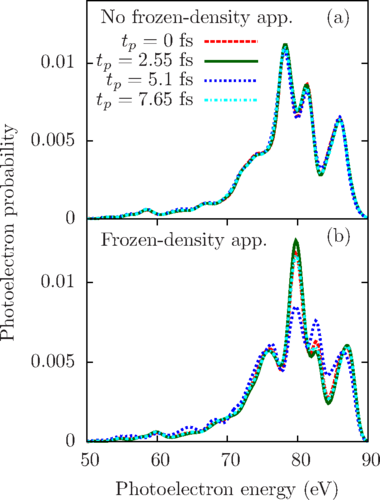}
\caption{{Time-resolved photoelectron spectra in the case of the probe-pulse duration being comparable with the shortest beating period of the electron density of indole (\textbf{a}) computed with Equation~(\ref{EqPE_nfrzd}) accurately treating electron dynamics during the probe and (\textbf{b}) computed with Equation~(\ref{EqPE}) neglecting electron dynamics during the probe.} Figure is reproduced with permission from Reference~\cite{Popova-GorelovaPRA16}. Copyright American Physical Society, 2016.
\label{SpectraFigAv}
}
\end{figure}

 The authors compared angle-averaged photoelectron spectra at four probe-pulse arrival times calculated with Equation~(\ref{EqPE_nfrzd}) taking into account the evolution of the electronic system during the action of the probe pulse shown in Figure~\ref{SpectraFigAv}a and with Equation~(\ref{EqPE}) assuming the frozen-density approximation shown in Figure~\ref{SpectraFigAv}b. The dramatical difference between Figure~\ref{SpectraFigAv}a and Figure~\ref{SpectraFigAv}b demonstrated that the frozen-density approximation broke down and provided completely incorrect results. Thus, they~demonstrated that if the probe-pulse duration is comparable with or longer than the shortest beating period of the electron density of a given system, the generated photoelectron spectra are not connected to its instantaneous electronic state as assumed in Equation~(\ref{EqPE}). At the same time, they~verified that a probe-pulse duration being ten times shorter than the oscillation period of the electron density is enough to satisfy the frozen-density approximation.

Let us consider the consequence of the evolution of the electronic system during the action of the probe pulse on time-resolved scattering patterns obtained by means of NRXS and RXS. The~corresponding relation taking into account the evolution of $\Psi(t)$ can be derived for NRXS from the general relation in Equation~(\ref{EqDSP_NRXS_Gen}) resulting in
\begin{align}
\frac{dP^{\NRXS}(\Q)}{d\Omega} =&\mathrm P_0^{\NRXS} \sum_F \int d^3 r_1\int d^3 r_2 e^{i\Q\cdot(\br_1-\br_2)}\label{EqDSP_NRXS_nfrzd} \\
&\qquad\times\sum_{I,K}\la\Psi_K(t_p)|\hat G_F^\NRXS(\br_1,\br_2)|\Psi_I(t_p)\ra e^{-\Omega_{FK}^2\tau_p^2/(8\ln2)} e^{-\Omega_{FI}^2\tau_p^2/(8\ln2)},\nonumber
\end{align}
 where $\Omega_{FI(FK)}=\omega_\i-\omega_\ks-E_F+E_{I(K)}$. In addition, such a relation for RXS results from the general relation in Equation~(\ref{EqDSP_RXS_Gen}), leading to
\begin{align}
\frac{dP^\RXS(\Q)}{d\Omega}=& \mathrm P_0^{\RXS}
\sum_{C_1,C_2}e^{i\mathbf Q\cdot(\mathbf R_{C_1}-\mathbf R_{C_2})}\sum_F \sum_{I,K}\la\Psi_K(t_p) |  \hat  G_{F,C_1,C_2}^\RXS |\Psi_I(t_p)\ra e^{-\Omega_{FK}^2\tau_p^2/(8\ln2)} e^{-\Omega_{FI}^2\tau_p^2/(8\ln2)}.\label{EqDSP_RXS_nfrzd}
\end{align}

In both cases, the relations are not connected to the electronic state at the time of measurement $\Psi(t_p)$ that cannot be singled out. Instead of that, all three relations in Equations~(\ref{EqPE}), (\ref{EqDSP_NRXS_nfrzd}) and (\ref{EqDSP_RXS_nfrzd}) include the electronic state $\Psi(t_p)$ convoluted with the amplitude of the electric field [{cf.~}Equation~(\ref{EqGaussianPulse})]. Thus, an assumption that a time-resolved signal obtained by a probe pulse not satisfying the frozen-density approximation depends on the electronic state simply averaged over the pulse duration, $\int_{t_p-\tau_p/2}^{t_p+\tau_p/2}dt\Psi(t)/\tau_p$, is incorrect.


Actually, Equations~(\ref{EqDSP_NRXS}), (\ref{EqDSP_RXS}) and (\ref{EqPE}) assuming the frozen-density approximation can be obtained from corresponding Equations~(\ref{EqPE_nfrzd})--(\ref{EqDSP_RXS_nfrzd}) by substituting the mean energy $\la E \ra$ of the superposition $\Psi(t_p)$ for $E_{I(K)}$ in factors $\Omega^\el_{FI(FK)}$ and $\Omega_{FI(FK)}$, correspondingly. This means that the frozen-density approximation is valid as long as the maximum energy difference of the eigenstates involved in the dynamics is negligible compared to the probe-pulse bandwidth.

Since the dependence of Equations~(\ref{EqDSP_NRXS_nfrzd}) and (\ref{EqDSP_RXS_nfrzd}) on the probe-pulse duration $\tau_p$ encoded in the factors $e^{-\Omega_{FI(FK)}^2\tau_p^2/(8\ln2)}$ is analogous to this dependence in the case of TRAPES in Equation~(\ref{EqPE_nfrzd}), it~is possible to transfer the conclusion of Reference~\cite{Popova-GorelovaPRA16} to the case of time-resolved X-ray scattering. Thus, if~the duration of an X-ray probe-pulse is about ten times shorter than the shortest beating period of the electron density, then time-resolved scattering patterns are connected to the electronic state $\Psi(t_p)$ and can be described by corresponding Equations~(\ref{EqDSP_NRXS}) and (\ref{EqDSP_RXS}). Whereas, if the duration of the X-ray probe pulse is comparable with the shortest beating period, then this connection is lost, and scattering patterns are determined by $\Psi(t_p)$ convoluted with the amplitude of the probe-pulse electric field as described by corresponding Equations~(\ref{EqDSP_NRXS_nfrzd}) and (\ref{EqDSP_RXS_nfrzd}).



\section{Discussion}
\label{SectionDiscussion}

We reviewed the theoretical framework based on QED developed for an accurate description of an interaction between a coherently evolving electronic system and an ultrashort light probe pulse. We~concentrated on its application to three techniques, namely, time-resolved NRXS, time-resolved RXS and TRARPES that can be employed for a measurement of electron dynamics in real space and real time  \cite{DixitPNAS12, DixitPRL13, DixitPRA14, BiggsJPhB14, BennettJChPh14, GrosserPRA17, KowalewskiStrDyn17, PopovaGorelovaPRB15_1, PopovaGorelovaPRB15_2, Popova-GorelovaPRA16}. It turned out that the corresponding theories and their results have much in common.

The outcomes of the theoretical analyses appeared to be quite counter-intuitive in all considered cases. Time- and momentum-resolved signals from a nonstationary electronic system obtained by means of the considered time-resolved techniques encode different information from their stationary analogues. X-ray scattering, which is determined by the electron density of a stationary object being measured, is not determined by the time-dependent electron density in a time-resolved measurement. The connection of scattering patterns to the electron density relay on elastic scattering dominating over inelastic scattering. In the case of an interaction of a light pulse with a nonstationary electronic system, the concept of elastic scattering is ambiguous, since it is extremely improbable that a final state after the interaction would be the same nonstationary electronic state. Moreover, such a situation would not be spectroscopically distinguishable from other possible inelastic scattering events. As a result, time-resolved scattering patterns encode spatial and temporal correlations substantially deviating from quantities encoded in stationary scattering patterns.

A similar consequence of a time-resolved measurement on photoelectron spectroscopy has been demonstrated. Contrary to an intuitive assumption that electron dynamics can be measured by means of time-dependent chemical shifts (i.e., by following temporal changes of electronic binding energies), time-resolved photoelectron spectra consist of a series of photoelectron peaks centered on time-independent positions. Electron dynamics is indeed encoded in time-dependent amplitudes of these peaks, which exhibit prominent temporal dependence in photoelectron angular distributions at fixed photoelectron energies. 

In all cases considered, time- and momentum-resolved signals,  i.e., scattering patterns in the case of NRXS and RXS, and photoelectron angular distributions in the case of TRARPES, do not follow instantaneous electron density. They depend on the phase of the electronic wave packet being measured and are not centrosymmetric with respect to the momentum (except for certain time points). It turns out that they have a quite similar dependence on the electronic state at the time of measurement $\Psi(t_p)$ and the spectral density of a probe pulse. In all cases, they are connected to a function, which can be represented as $\la \Psi(t_p)|\hat H(\br_1,\br_2)|\Psi(t_p)\ra$, where an operator $\hat H(\br_1,\br_2)$ includes a sum over final states of corresponding transition amplitudes weighted by a function determined by the spectral density of a probe pulse. The exact form of the operator $\hat H(\br_1,\br_2)$ depends on the technique by which the signal was obtained.

There are two contributions to the time- and momentum-resolved signals. The first contribution is centrosymmetric with respect to the momentum and is determined by a function connected to the real parts of electron density matrix elements. This means that the temporal evolution of this function correlates with the time-dependent electron density and charge distributions. The second contribution is an odd function with respect to the momentum and is determined by a function connected to the imaginary parts of electron density matrix elements. Its temporal dependence correlates with electron currents. Thus, the fact that the signals are not centrosymmetric reflects that electrons are moving. The two contributions can be disentangled by performing the Fourier transform from momentum space to real space. Specific connections of these functions to certain time-dependent quantities have to be identified for a given electronic system and technique. For instance, such an analysis has been performed for the case of ultrafast RXS in References~\cite{PopovaGorelovaPRB15_1, PopovaGorelovaPRB15_2}, where a method to measure interatomic electron currents was introduced.

The connections of the time-resolved signals to the electronic state $\Psi(t_p) $ hold as long as the bandwidth of a probe pulse is much larger than the maximum energy difference between electronic states involved in the dynamics. This condition is satisfied for a probe-pulse duration being ten times shorter than the shortest beating period of the electron density. If the probe-pulse duration is comparable with this period, when the time-resolved signals are determined by the convolution of the electronic state with the electric-field amplitude of the probe pulse, but not by its temporal average over the duration of the probe-pulse.


\vspace{6pt}


\acknowledgments{The author acknowledges valuable discussions with Robin Santra.}

\conflictsofinterest{The author declares no conflict of interest.}

\abbreviations{The following abbreviations are used in this manuscript:\\

\noindent 
\begin{tabular}{@{}ll}
QED & quantum electrodynamics\\
NRXS & nonresonant X-ray scattering\\
RXS & resonant X-ray scattering\\
TRARPES & time- and angle-resolved photoelectron spectroscopy \\
XUV & extreme ultraviolet\\
DSP & differential scattering probability \\
HOMO & highest occupied molecular orbital
\end{tabular}}

\symbols{The following symbols are used in this manuscript:\\
\noindent 
\begin{longtabu}{@{}lll}
$\Psi(t)$ & state of a coherently evolving electronic system \\
$\Phi_I$, $\Phi_K$ & electronic states comprising the coherent superposition \\
$C_I$, $C_K$ & complex time-independent coefficients of the coherent superposition\\
$E_I$, $E_K$ & energies of the electronic states comprising the coherent superposition\\
$\la E\ra$ & mean energy of the coherent superposition \\
$\Phi_F$ & final state\\
$E_F$ & final-state energy\\
$\rho$ & electron density\\
$\hat\rho^m$ & density matrix of an electronic system\\
$\hat\rho_f$ & total density matrix of the matter and the electromagnetic field\\
$\mathbf j$ & electron current density \\
$\psi$ & electron annihilation field operator\\
$T$ & oscillation period of an electronic wavepacket with two eigenstates \\
$t_0$ & time, when the coherent superposition was created\\
$t_p$ & time of measurement\\
$\tau_p$ & probe-pulse duration\\
$t_f$ & time after the action of the probe pulse\\
$\k$ & photon momentum\\
$p$ & index referring to a photon polarization\\
$\hat a_{\k,p}$ & annihilation operator of the photon in the $\k$, $p$ mode\\
$\omega_{\k}$ & photon energy\\
$c$ & speed of light\\
$G^{(1)}$ & first-order radiation field correlation function\\
$\mathcal E$ & electric-field amplitude of a probe pulse \\
$I_0$ & probe-pulse peak intensity\\
$\mathbf r_0$ & position of an object\\
$\{n\}$, $\{\widetilde n\}$ & complete sets that specify the number of photons in all initially occupied field modes \\
$\rho ^X_{\{n\},\{\widetilde n\}}$ & distribution of all occupied field modes associated with the probe pulse\\
$\omega_\i$ & probe-pulse photon energy\\
$\boldsymbol\epsilon_{\i}$ & probe-pulse polarization\\
$\boldsymbol\epsilon_{s}$ & polarization of a scattered photon\\
$\Q$ & scattering vector\\
$\omega_{\k_\s}$ & scattering energy\\
$\k_{\s}$ & momentum of a scattered photon\\
$\mathbf A$ & vector potential\\
$\mathbf p$ & canonical momentum of an electron\\
$V$ & quantization volume \\
$\hat H$ & total Hamiltonian of the whole system, matter and light \\
$\hat H_{\text{m}}$ & Hamiltonian of the electronic system \\
$\hat H_{\text{int}}$ & interaction Hamiltonian between the matter and the electromagnetic field\\
$\hat H_{\text{int}}^{(1)}$ & interaction Hamiltonian determined by the $\mathbf A\cdot\mathbf p$ term\vspace{2pt}\\
$\hat H_{\text{int}}^{(2)}$ & interaction Hamiltonian determined by the $\mathbf A^2$ term \vspace{2pt}\\
$P^{\text{XS}}$ & probability of X-ray scattering\\
${dP}/{d\Omega}$ & differential scattering probability\\
${dP^{\text{st}}}/{d\Omega}$ & differential scattering probability according to the quasistationary treatment\\
$\mathbf R_C$ & position of an atom $C$\\
$f_C$ & scattering amplitude of an atom $C$\\
$C$, $C_1$, $C_2$ & atomic indices\\
$\Phi_{J_C}$ & intermediate state with an electron hole in a core shell of an atom $C$ in the RXS process\\
$\Gamma_{J_C}$ & decay width of the intermediate state $J_C$ \\
$E_{J_C}$ & energy of the intermediate state $J_C$ \\
$j_{C_1C_2}$ & interatomic electron current between atoms $C_1$ and $C_2$\\
$W(\omega_{\k_\s})$ & function representing photon-detector acceptance range\\
$ P^\PE$ & photoelectron probability\\
$\Q_\el$ & photoelectron momentum\\
$\sigma$ & photoelectron spin\\
$\phi_{\sigma e}$ & photoelectron wave function \\
$\varepsilon_e$ & photoelectron energy\\
$\hat c_{\Q_\el,\sigma}$ & annihilation operator of an electron with momentum $\Q_\el$ and spin $\sigma$\\
 $N_{\el}$ & number of electrons in a system before the action of a photoionizing probe pulse\\
 $\mathcal F$& Fourier transform from momentum space to real space\\
$\hat \U$ & time-evolution operator\\
$|\Psi_{\{n\}},t_f\ra$ & wave function being an entangled state of the electronic and photonic states\\
$\Omega_F$ & factor equal to $\omega_\i-\omega_{\k_\s}-E_F+\la E\ra$ \\
$\Omega_F^\el$ & factor equal to $\omega_\i-E_F^{N_{\el}-1}+\la E \ra$ \\
$\Omega_{FI(FK)}$ & factors equal to $\omega_\i-\omega_\ks-E_F+E_{I(K)}$ \\
$\hat G_F^\NRXS$ & operator defined in Equation~(\ref{EqGF_NRXS})\vspace{2pt}\\
$\hat  G^\RXS_{F,C_1,C_2}$ & operator defined in Equation~(\ref{EqGF_RXS})\vspace{2pt}\\
$\hat G_F^\PE$ & operator defined in Equation~(\ref{EqGF_PE})\\
$\mathrm P_0^{\NRXS}$ & factor equal to $I_0\tau_p^2\sum_{p_\s}\bigl|(\boldsymbol\epsilon_{\i}\cdot \boldsymbol\epsilon^*_{\s})\bigr|^2/(4\ln2\,\omega_{\i}^2c^3)$\\
$\mathrm P_0^{\RXS}$ & factor equal to $\tau_p^2I_0/(4\ln2c^4)$\\
$\mathrm P_0^{\PE}$ & factor equal to $I_0\tau_p^2/(8\pi\ln2\omega_\i^2c)$\\
 $\mathcal R^\NRXS$, $\mathcal R^\RXS$, $\mathcal R^\PE$ & functions determining centrosymmetric contribution to time- and momentum-resolved\\ & NRXS, RXS and PE signals, correspondingly\\
 $\mathcal J^\NRXS$, $\mathcal J^\RXS$, $\mathcal J^\PE$ & functions determining noncentrosymmetric contribution to time- and momentum-resolved\\ & NRXS, RXS and PE signals, correspondingly\\
  $A[\phi_F^D(t_p)] *s$ & convolution of the autocorrelation function of the time-dependent Dyson orbital $\phi_F^D$ \\ & with $\operatorname{sinc}(Q_0|\mathbf r|)$\\
\end{longtabu}}


\reftitle{References}




\end{document}